\documentclass[structabstract]{aa}  
\usepackage{graphicx}
\usepackage{longtable}
\usepackage{lscape}
\usepackage{txfonts}
\begin{document}
\title{Constraining the population of isolated massive stars within 
the Central Molecular Zone.
\thanks{Based on observations made at the European Southern Observatory, 
Paranal, Chile under programmes ESO 093.D-0168} 
}
\author{J.~S.~Clark\inst{1}
\and L.~R.~Patrick\inst{2,3}
\and F.~Najarro\inst{4}
\and C.~J.~Evans\inst{5}
\and M.~Lohr\inst{1}
}
\institute{
$^1$Department of Physics and Astronomy, The Open 
University, Walton Hall, Milton Keynes, MK7 6AA, United Kingdom\\
$^2$Instituto de Astrof\'{\i}sica de Canarias, E-38205 La Laguna, Tenerife, Spain\\ 
$^3$DFISTS, EPS, Universidad de Alicante, Carretera
San Vicente del Raspeig s/n, E-03690 San Vicente del Raspeig, Spain\\
$^4$Departamento de Astrof\'{\i}sica, Centro de Astrobiolog\'{\i}a, 
(CSIC-INTA), Ctra. Torrej\'on a Ajalvir, km 4,  28850 Torrej\'on de Ardoz, 
Madrid, Spain\\
$^5$UK Astronomy Technology Centre, Royal Observatory Edinburgh, Blackford Hill, 
Edinburgh EH9 3HJ, UK
}

   \abstract{Many galaxies host pronounced circumnuclear starbursts, fuelled by infalling gas. 
Such activity is expected to drive the secular evolution of the nucleus and generate super winds, while the intense radiation fields and 
extreme gas and cosmic ray densities present may act to modify the outcome of star formation with respect to more quiescent galactic regions.} 
{The centre of the Milky Way is the only example of this phenomenon where, by virtue of its proximity,  individual stars may be resolved. Previous studies reveal that it hosts a rich population of very massive stars; these are  located within three clusters, with an additional cohort  dispersed throughout the Central Molecular Zone. In this paper we investigate the size and composition  of the latter contingent.}
{We utilised the VLT+KMOS  to obtain homogeneous, high signal-to-noise ratio observations of known and candidate massive stars suitable for spectral classification and quantitative analysis. }
{Including previously identified examples, we find a total of  83 isolated massive stars within the Galactic Centre, which are strongly biased towards objects supporting  powerful stellar winds and/or extensive circumstellar envelopes. No further stellar clusters, or their tidally stripped remnants, were identified, although an apparent stellar overdensity is found to be coincident with the Sgr B1 star forming region.}
{The cohort of isolated massive stars within the Central Molecular Zone is directly comparable in size to that of the known clusters and, due to observational biases, is likely highly incomplete at this time.  Combining both populations yields $\gtrsim320$ spectroscopically classified stars that would be  expected to undergo core collapse within the next $\sim20$Myr. Given that this is almost certainly a substantial  underestimate of the true number, the population of massive stars associated with the Galactic Centre appears unprecedented amongst star formation complexes within the Milky Way, and it appears probable that they  play a substantial role in the energetics and evolution of the nuclear region.}


\maketitle

\section{Introduction}

While many galaxies host prominent (circum-)nuclear starbursts, the physics governing their formation and subsequent contribution to the wider galactic ecology and energy budget is currently opaque due to  our inability to resolve individual stars in such environments at extragalactic distances.  Indeed there is only one example of this phenomenon - the central region of our own Galaxy - where this is currently possible. Multiwavelength observations have revealed that physical conditions in the Galactic Centre (GC) are particularly  extreme with respect to the disc, bearing close resemblance to  those anticipated for high redshift starburst galaxies (Kruijssen \& Longmore \cite{dkl}), hence the hope that the GC will act as a template for such objects.  Specifically, the mean temperature, density, pressure, and velocity dispersion of molecular material; the magnetic field strength; the cosmic ray density; and the ionisation rate are significantly greater than those found in the Galactic disc, in some cases by orders of magnitude. As such, one might anticipate that processes such as star formation proceed in a different manner than in more quiescent regions of the Milky Way.

It is important to determine if  this is the case. Encompassing the inner $\sim500$pc of the GC, the Central Molecular Zone (CMZ) contains up to $\sim10$\% of the molecular mass of the Galaxy ($\sim2-6\times10^7M_{\odot}$; Morris \& Serabyn \cite{MS}), which in turn fuels the most extreme star forming region  within the Milky Way. Observations from sub-mm to radio wavelengths suggest this activity is occurring  at multiple locations within the H\,{\sc ii} regions populating the GC;
within $\sim6$pc of Sgr A$^*$ (Yusef-Zadeh et al. \cite{yz10}), Sgr B1+2  
(e.g. Ginsburg et al. \cite{ginsburg18a}, Hankins et al. \cite{hankins}), and  Sgr C 
(Kendrew et al. \cite{kendrew}), as well as the dust ridge linking Sgr A to Sgr B
(Immer et al. \cite{immer12b}) and  mid-IR hotspots associated with ionised gas between Sgr A and C (Hankins et al. \cite{hankins}).
Additional mid-IR surveys have identified numerous isolated point sources distributed throughout the CMZ with properties that are consistent with young stellar objects (Yusef-Zadeh et al. \cite{yz09}, An et al. \cite{an}, Immer et al. \cite{immer12a}). A compact aggregate of such sources $\sim8$arcmin north of the Sgr C H\,{\sc ii} region is suggestive of cluster  formation (Yusef Zadeh et al. \cite{yz09}); a conclusion that is buttressed by sub-mm observations of the Sgr B2 region, which imply a combination of clustered and distributed star formation (Ginsburg et al.   \cite{ginsburg18a}, Ginsburg \& Kruijssen \cite{ginsburg18b}).

However,  despite this vigorous  activity and presence of copious molecular material, estimates of the star formation rate for the CMZ suggest that it is at least an order of magnitude lower than expected based on observations of nearby regions (Longmore et al. \cite{longmore13}). The physical cause of this discrepancy is uncertain  (Barnes et al. \cite{barnes}), but it raises  the possibility that the  resultant stellar population(s) may also show an environmental dependance, possibly characterised by an anomalous initial mass function (IMF). As a consequence, much effort has been expended in attempts to characterise the young stellar population within the CMZ, with 
near-IR observations revealing a rich population of massive stars located within   clusters - the Arches, Quintuplet, and Galactic Centre (e.g. Figer et al. \cite{figer99}, Paumard et al. \cite{paumard}) - and distributed throughout the CMZ in apparent isolation (Cotera et al. \cite{cotera96}, \cite{cotera99}, Muno et al. \cite{muno06}, Mauerhan et al. \cite{mauerhan07}, \cite{mauerhan10a}, \cite{mauerhan10c}, Dong et al. \cite{dong15}).

Beyond constraining the star formation physics operating within the extreme conditions of the GC, a determination of the properties of this stellar cohort is of considerable importance for a number of  other astrophysical topics. With a subset born with masses $M_{\rm init}>100M_{\odot}$ (Lohr et al. \cite{lohr}), they provide vital observational data on the lifecycle of the most massive stars that form in the local Universe, up to and including the point of core-collapse. Constraining a robust evolutionary scheme for such stars is essential  if we are to predict both the nature and  production rate of relativistic remnants from this population, noting that a rich cohort  of young neutron stars and black holes deriving from such stars  appears present within the GC (Deneva et al. \cite{deneva}, Kennea et al. \cite{kennea}, Hailey et al. \cite{hailey}). 

It is anticipated  that massive stars also play an important role in driving the evolution of the GC via the feedback of ionising radiation, mechanical energy and chemically enriched material. Of particular interest is their role in shaping the emergent high energy spectrum of the GC, which  recent observations suggest extends from soft X-rays ($kT\sim1-10$keV; Ponti et al. \cite{ponti}) through  to very high energy $\gamma$-rays ($kT > 100$GeV; Aharonian et al. \cite{aharonian06}). It appears likely that the diffuse,
 low energy X-ray emission arises from a combination of unresolved low mass point sources (pre-MS stars and Cataclysmic Variables), the winds of massive stars and their  supernova (SN) endpoints - acting both individually and in concert in massive clusters such as Wd1 (Muno et al. \cite{muno06b}) - and pulsar wind nebulae (Ponti et al. \cite{ponti}). 

The  $\gamma$-ray component is thought to derive from the cosmic rays  that permeate the GC (Aharonian et al. \cite{aharonian06}, H.E.S.S. collaboration \cite{HESS16}, \cite{HESS18}). An exceptional  cosmic ray density  may be inferred by the abundance of H$_3^+$ (produced via the ionisation of H$_2$) and has been suggested to play an important role in regulating the temperature of the warm molecular material that suffuses the CMZ (Le Petit et al. \cite{lepetit}, Oka et al. \cite{oka}). 
Plausible sources for the production of cosmic rays are the supermassive black hole Sgr A$^*$ and massive stars - the latter via the interaction between their  winds, cluster driven outflows  and supernovae  (e.g. Aharonian et al. \cite{aharonian18}, Bednarek et al. \cite{bed}, Bykov et al. \cite{bykov}, Cesarsky \& Montmerle \cite{CM}). Unfortunately, in the absence of a full stellar census the relative contributions of these channels  is currently uncertain.  

Nevertheless both physical agents  have been implicated in the initiation of  mass outflows - thought to be  driven by a combination of cosmic ray and thermal gas pressure (cf. Everett et al. \cite{everett}, Yusef-Zadeh \& Wardle \cite{yz19}) -  that   originate in the GC. These range in size from the $\sim15$pc radio and X-ray lobes (Morris et al. \cite{morris03}, Zhao et al. \cite{zhao}) through to the order of magnitude larger  bipolar radio bubbles and X-ray chimney (Heywood et al. \cite{heywood} and  Ponti et al. \cite{ponti19}, respectively), and ultimately the $\sim50$kpc Fermi bubbles (Su et al. \cite{su}). An intriguing possibility is that the removal of material from the CMZ via such winds may help quench star formation, leading to the low rate currently observed.

The preceding discussion leads to the conclusion that a more complete understanding of the population of massive stars within the GC is extremely timely and well motivated for a multitude of reasons.
With the Galactic Centre cluster cohort well constrained (Paumard et al. \cite{paumard}, Bartko et al. \cite{bartko}) previous papers in this series have focused on a reappraisal of the Arches and Quintuplet clusters (Clark et al. \cite{clark18a},
\cite{clark18b}, \cite{clark19a}). In this work  we concentrate on the apparently isolated massive stellar component distributed throughout the GC, utilising extant surveys (Sect. 4.1) to compile a target list which we observed with the K-band multi-object spectrograph (KMOS)  mounted on UT1 of the Very Large Telescope (VLT). The manuscript is ordered as follows. Sect. 2 details data acquisition and reduction and the classification criteria employed to characterise the resultant spectra. Sect. 3 provides a detailed breakdown of the resultant dataset by spectral subtype,  while we discuss survey completeness, the distribution of massive stars across the CMZ and construct a complete stellar census for the CMZ in Sect. 4. Finally we summarise our findings and highlight future prospects to advance these research lines in Sect. 5. 

\section{Data acquisition, reduction,  and classification}

\subsection{Acquisition and reduction}

The VLT-KMOS (Sharples et al. \cite{sharples}) data for this paper
were obtained under  ESO programme 093.D-0306 (PI: Clark),
with observations made between 2014 August 02-13. KMOS is a multi-object,
integral field spectrograph, which has 24 configurable integral field units
(IFUs) positioned within a 6.7 arcminute field of view.
The spectral resolution of the observations is a function of rotator 
angles and the IFUs used (e.g. Patrick et al. \cite{patrick15}),
 varying between ${\Delta}{\lambda}/{\lambda}\sim3895 - 4600$. Each 
observing block consisted of 12$\times$30\,s exposures in an ABA
observing pattern, where the first observation of each field used the more
rigorous 24-arm telluric standard star approach and all subsequent
observations of the same field used the standard 3-arm telluric approach.
The standard stars used for these observations were HIP\,84846 (A0V),
HIP\,91137 (A0V), and HIP\,3820 (B8V).

The data reduction
methodology is identical to that of the KMOS data presented in Clark et al.
(\cite{clark18b}).
Science and standard star observations were calibrated, reconstructed and
combined using the KMOS/esorex pipeline (Davies et al. \cite{kmos}), employing the standard set of
calibrations delivered by the telescope. Clark et al. (\cite{clark18b}) detail this
procedure and discuss the modifications made to the standard processes.

In the K-band,  telluric correction is a
fundamentally important part of the data reduction process. Since 
the majority of the useful diagnostic lines for these targets lie in 
regions of the K-band that are 
highly contaminated by telluric absorption, we
implemented a rigorous correction routine adapted from Patrick 
et al. (\cite{patrick15}, \cite{patrick17}) and further detailed in Clark et
al. (\cite{clark18b}).  Given the intrinsic
shape of the telluric spectrum, the continuum placement is vital to
accurately recover the shape of the science spectrum. This is typically done
empirically, by selecting multiple continuum points from the science and
standard star spectra throughout the entire spectral range and is highly
non-linear.

For targets with particularly broad spectral features, such as the WN and WC stars, 
identification of the continuum is a complicated process.
 This problem is compounded when broad emission  features coincide
with strong telluric absorption, as seen for the He\,{\sc ii} 2.0379$\mu$m and 2.3799$\mu$m
and He\,{\sc i} 2.059$\mu$m features in the WN5-7 stars. For such stars continuum placement was guided 
by comparison with a combination of published  spectra - in particular 
that of qF353E (WN6; Steinke et al. \cite{steinke}) - and synthetic
examples computed with from the CMFGEN code (Hillier \& Miller \cite{hillier98}, \cite{hillier99}).

\subsection{Spectral classification}

A number of publications have been dedicated to the classification of post-MS massive 
stars in the near-IR window: specifically O  stars (Hanson et al. \cite{hanson96}, \cite{hanson05}), B-hypergiants 
(Clark et al. \cite{clark12}, \cite{clark18b}),
luminous blue variables (LBVs; Morris et al. \cite{morris}, Clark et al. \cite{clark11})  supergiant B[e] stars (sgB[e];
Oksala et al. \cite{oksala}) and Wolf-Rayets (WRs; Figer et al. \cite{figer97}, Crowther et al. 
\cite{crowther06}, Crowther \& Walborn \cite{crowther11} and Rosslowe 
\& Crowther \cite{rosslowe}).

We have employed  - and expanded upon - these classification criteria in our study 
of the Arches and Quintuplet (Clark et al. \cite{clark18a}, 
\cite{clark18b}). We follow an identical methodology here, referring the reader to these works 
for details beyond those summarised in the relevant sections below. As in previous works, 
given the uncertainty in the parameterisation of the spatially inhomogeneous interstellar 
extinction along sightlines towards the GC, we prioritise spectral rather than photometric data. 
However given that we are unable to utilise cluster membership to locate target stars within the GC, we are forced 
to employ the photometric datasets and analysis of Dong et al. (\cite{dong12}) to identify likely foreground 
interlopers. In doing so we choose to only make use of  ground based photometry to avoid issues of calibrating such data with space-based 
observations, given the significant issues accounting for convolving very different filter responses for intrinsically red
 photometric sources (cf. Dong et al. \cite{dong12}). 

For candidate massive stars previously identified in  the literature but without KMOS observations we used published spectra to  reappraise their classifications in light of this methodology. Where stars are reclassified on this basis, if we were unable to obtain  the relevant spectra we  provide the appropriate figure number in addition to the formal reference in the following discussion.

In the remaining sections for conciseness, we abbreviate the [DWC2011]{\em xxx} designation for stars in the primary list of Pa$\alpha$ emitters presented in Dong et al. (\cite{dong11}) to a simple P{\em xxx}. No recognised nomenclature exists for those stars derived from the secondary list of Pa$\alpha$ emitters from this paper; hence we choose to designate these simply as S{\em xxx} sources.

\section{Results}

Observations were made of a total of 82 candidate and confirmed massive stars derived from the list of Pa$\alpha$ excess sources of Dong
et al. (\cite{dong11}) and other literatures sources (Mauerhan et al. \cite{mauerhan07}, \cite{mauerhan10a}, \cite{mauerhan10c}).
Details of each target, including previous and new classifications are provided in Table 1.
 The sample contains  a diverse group of objects including pre- and post-main sequence massive stars as well as a large number of foreground high- and low-mass interlopers. Below we break down this population by spectral sub-types and location along the sightline to the GC, including discussion of relevant examples not included in our sample in order to provide the basis for the construction of a comprehensive stellar census.

\subsection{OB  supergiants}

Spectra of OB supergiants are presented in Fig. 1. We are able to identify three new candidates; S73, S152, and P95. The first two stars  are clearly  mid-O supergiants given the presence of He\,{\sc ii} 2.189$\mu$m absorption and C\,{\sc iv} 2.069+2.078$\mu$m emission. The lack of a 
pronounced He\,{\sc i} $\sim2.112\mu$m absorption feature in the He\,{\sc i}+N\,{\sc iii}+C\,{\sc iii} blend of either star indicates  comparatively early spectral types (O4-5), while the relatively weak   Br$\gamma$ photospheric line signals significant mass loss, though not sufficient to drive the line into emission as is seen in  hypergiants (Fig. 2). 
Three additional mid-O supergiants - CXOGC J174628.2-283920, 174703.1-285354, and 174725.3-282523 - were observed by  
Mauerhan et al. (\cite{mauerhan10c}). These were not observed with KMOS, but we include their published  classifications in Table 1 in order to compile a comprehensive census of massive stars in the GC region.

Moving to later spectral sub-types and the presence of narrow photospheric absorption in He\,{\sc i} 2.059, 2.112,  and 2.161$\mu$m, Br$\gamma$ and He\,{\sc ii} 2.189$\mu$m indicates that P95 is a new $\sim$O9 supergiant via close similarity to the spectral template HD154368 (Hanson et al. \cite{hanson05}); however the narrower Br$\gamma$ profile and stronger He\,{\sc i} 2.112$\mu$m emission suggests a stronger wind than this classification standard. 
Given the S/N of our spectrum of CXOGC J174537.3-285354 around 2.19$\mu$m, we may not improve on the previous 
O9-B0Ia  classification, nor reassess the nature of P50 from the spectrum presented in  Mauerhan et al. (\cite{mauerhan10c}). Geballe et al. 
(\cite{geballe}; their Fig. 6) identify 2MASS J17444501-2919307 as  B2-3 Ia$^+$; we prefer a slightly more conservative B0-3 Ia classification. 
Finally  De Witt et al. (\cite{dewitt}) propose a generic O star classification for XID 947;  given the low S/N of the published spectrum we are unable to improve on this.

\begin{figure*}
\includegraphics[width=11cm,angle=-90]{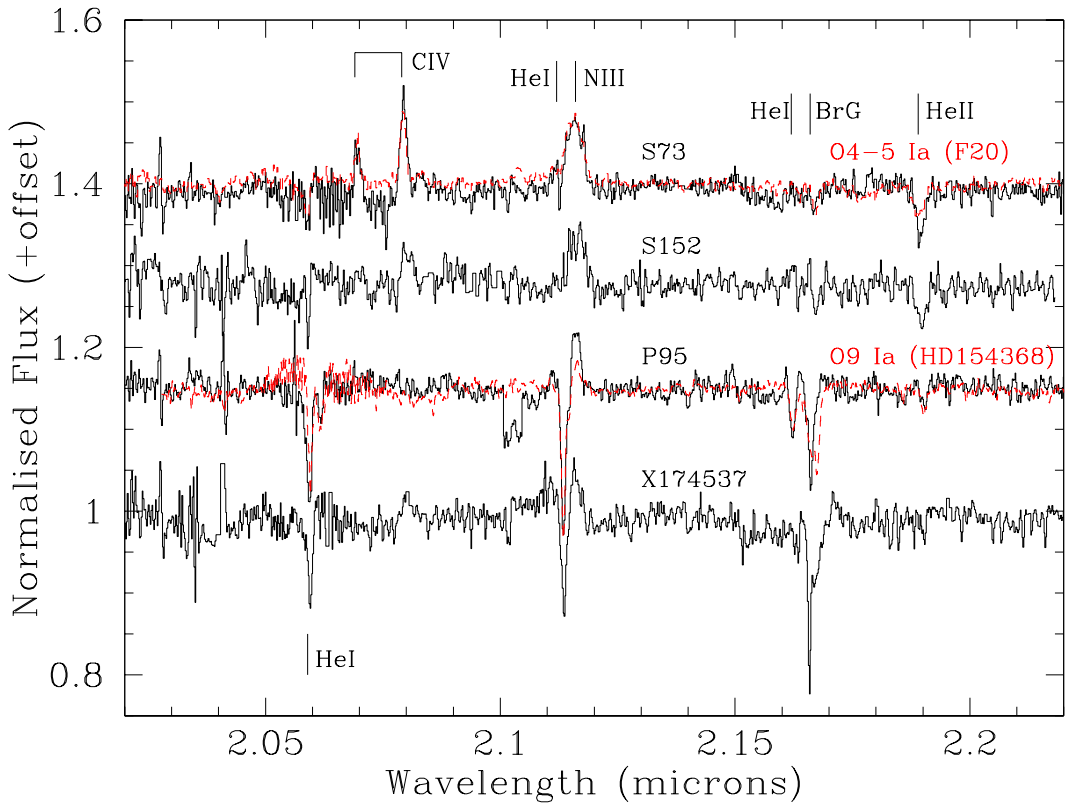}
\caption{Montage of spectra of OB supergiants. Template spectra with classifications from Hanson et al. (\cite{hanson05}) 
and Clark et al. (\cite{clark18a}) overplotted in red. It is important to note that hot pixels around 2.080$\mu$m - nearly coincident with the C\,{\sc iv} 2.079$\mu$m lines - in the spectrum of CXOGC J174537.3-285354 were artificially removed; we suspect the anomalously narrow component of the Br$\gamma$ photospheric profile is also spurious. Likewise the broad absorption feature centred on $\sim2.1\mu$m in the spectrum of
P95 is also artificial.}
\end{figure*}

\subsection{O hypergiants and WN7-9ha stars}

Consideration of the spectra of members of the Arches reveals the close evolutionary and morphological similarities between 
$\sim$O4-8 hypergiants and WN7-9ha Wolf-Rayets (Martins et al. \cite{martins08}, Clark et al. \cite{clark18a}). The former 
are delineated by systematically weaker Br$\gamma$ emission and 
a P Cygni absorption component in the $\sim2.11\mu$m He\,{\sc i}+C\,{\sc iii}+N\,{\sc iii}+O\,{\sc iii} emission blend at later (O6-8) spectral subtypes. In contrast no absorption component is present in the $\sim2.11\mu$m feature of any WNLha star, while He\,{\sc ii} 2.189$\mu$m is in emission in the earlier ($<$WN7-8) spectral subtypes.

As a consequence we discuss both types of star here; presenting spectra of three new examples in Figs. 2 and 3. Of these P15 and S131 are clearly new O hypergiants by virtue of broad, pronounced Br$\gamma$ emission and  He\,{\sc ii} 2.189$\mu$m absorption.  Strong C\,{\sc iv} emission and a lack of He\,{\sc i} 2.112$\mu$m absorption indicates that P15 is an early O4-5 Ia$^+$ star (Clark et al. \cite{clark18a}, Hanson et al. \cite{hanson05}). Conversely the presence of He\,{\sc i} 2.112$\mu$m absorption and weak  C\,{\sc iv} emission in the spectrum of S131
indicates a later (O7-8 Ia$^+$) subtype; we examine the possible causes of the  double peaked Br$\gamma$ emission line profile below.

Of those stars with previous classifications,  the detection of Br$\gamma$ emission in P36 and 114 marks them out as hypergiants
rather than supergiants and of early (O4-5) spectral type, given the lack of He\,{\sc i} 2.112$\mu$m absorption and consequent 
similarity to the O4-5 Ia$^+$ Arches star F27 (Fig. 2). The Br$\gamma$ emission line in the spectrum of  P36 shows a central reversal, which is also present in 
the O4 Ia$^+$ spectroscopic template HD15570 and  other hypergiants considered here (see below); unfortunately the presence of a narrow emission component of uncertain  origin prevents  interpretation of the corresponding line profile  of P114.
The similarity of P100 \& 107 to  Arches F10  suggests a revision to  slightly later spectral subtypes (O4-6 to O7-8; Fig. 2);  the strength of Br$\gamma$ emission in both stars suggest they are close to transitioning to a WNLha evolutionary phase. In terms of the strength of C\,{\sc iv} emission P97 is intermediate between these stars and  P15; we assign  an O6-7 subtype by comparison to Arches F15 (Fig. 2). 

The similarity  of P23 to P97 suggests  a comparable classification, although the former demonstrates stronger Br$\gamma$ 
emission, suggesting it too is close to becoming a WNLha star. As with S131 its Br$\gamma$ profile 
is strongly double peaked, while C\,{\sc iv} 2.079$\mu$m and the emission component of the 
$\sim2.11\mu$m He\,{\sc i}+C\,{\sc iii}+N\,{\sc iii}+O\,{\sc iii} blend are unexpectedly broad. 
Regarding the Br$\gamma$ line, the only comparator we are aware of is the Quintuplet member LHO 001 
which is spectroscopically variable and  demonstrates a similarly double peaked profile at some epochs. 
Clark et al. (\cite{clark18b}) suggest that LHO 001 is a massive
binary system and such an explanation is also attractive for P23 as well; it is not obvious that a 
physically justifiable combination of He-abundance, mass-loss rate, wind clumping factor and velocity 
field for a single star can replicate the Br$\gamma$ line profile observed.

The Br$\gamma$ profile of P75 also appears double peaked, although the blue peak is less pronounced than 
in  P23, being more comparable to S131. Other notable features include strong, broad emission 
with hints of substructure  in the  $\sim2.11\mu$m He\,{\sc i}+C\,{\sc iii}+N\,{\sc iii}+O\,{\sc iii} 
blend, an absence of C\,{\sc iv} emission and, uniquely, weak N\,{\sc iii} 2.103$\mu$m emission.
The latter two  observational features are characteristic of the WN8-9ha stars rather than the  mid-O hypergiants 
within the Arches (Clark et al. \cite{clark18a}), although the reverse is true for the  He\,{\sc i} 
2.112$\mu$m absorption component also exhibited by P75. We suggest this unique hybrid morphology is due to 
strong helium and nitrogen enhancement (with the former yielding pronounced He\,{\sc i} 2.161$\mu$m 
emission in the blue wing of Br$\gamma$) and C depletion with respect to normal mid-O hypergiants as the star 
enters the WNLha phase. As such we revise the classification of P75 to WN9ha/O6-7 Ia$^+$; noting that
further multi-epoch observations and quantitative analysis are required to confirm the nature of stars such as P23, P75 and S131, which demonstrate 
double peaked Br$\gamma$ profiles.

Next we turn to the WNLha stars (Fig. 3).
Comparison of S120 to Arches F16 suggests that it is a new WN8-9ha star; the strength of Br$\gamma$ emission 
and lack of He\,{\sc i} 2.112$\mu$m absorption distinguishing it from an O hypergiant, while He\,{\sc ii} 2.189$\mu$m fully in absorption 
suggests a late spectral sub-type. Nevertheless there are incongruities; the Br$\gamma$ line and the He\,{\sc i}+C\,{\sc iii}+N\,{\sc iii}+O\,{\sc iii}
 $\sim2.11\mu$m emission blend both appear anomalously narrow in comparison to other isolated examples and the cohort within the Arches 
cluster (Clark et al. \cite{clark18a}). Moreover, the emission features exhibit a significant displacement from their rest wavelengths 
($\Delta RV{\gtrsim}100$kms$^{-1}$); possibly indicative of binary reflex motion or a runaway nature.

Assigned  a generic O If$^+$ classification (Muno et al. \cite{muno06}, Dong et al.  \cite{dong15}) P35 is of particular interest since it is spatially coincident with the H\,{\sc ii} region H2, which  Dong et al. (\cite{dong17}) associates with an apparent overdensity of bright stars. Inspection of our spectrum reveals exceptionally strong, narrow and asymmetric He\,{\sc i} 2.059$\mu$m and  Br$\gamma$ emission, the latter with a rather broad base (Fig. 3). Such a morphology is not characteristic of  O super-/hypergiants or WNLha stars. Conversely, weak C\,{\sc iv} 2.079$\mu$m emission, He\,{\sc ii} 2.189$\mu$m absorption and a strong 
broad pure emission profile in the He\,{\sc i}+C\,{\sc iii}+N\,{\sc iii}+O\,{\sc iii} $\sim2.11\mu$m blend is reminiscent of early-mid O hypergiants and weak-lined WN8-9ha stars. We therefore assign such a classification to P35, assuming that there is significant contamination of the He\,{\sc i} 2.059$\mu$m and  Br$\gamma$ profiles by nebular emission from the H2 H\,{\sc ii} region - as suggested by the inflection in the red flank of both lines. Finally, as discussed in Clark et al. (\cite{clark19a}) P96 closely resembles the WN7-8ha Arches member F4 (Table 1); the  
resolution  and S/N of the published spectra of the remaining candidates (cf. Mauerhan et al. \cite{mauerhan10a}, \cite{mauerhan10c}) being  insufficient to allow any further refinements to current  classifications.

Unlike the OB supergiants there exists sufficient quasi-homogeneous ground based photometry (Mauerhan et al. 
\cite{mauerhan10a}, Dong et al. \cite{dong11}) to  enable a comparison of the properties of isolated WN7-9ha 
and O hypergiants to those in the Arches cluster. Fig. 4 indicates that such stars within the Arches exhibit 
a range in  both $K-$band ($\sim9.5-11$) and $(H-K)$ colour indice ($\sim1.4-2.2)$. This is likely due to a combination of variations in both intrinsic (luminosity, temperature/bolometric correction and continuum emission from the stellar wind) and extrinsic properties (binarity, differential stellar reddening); 
indeed Lohr et al. (\cite{lohr}) find [FGR2002] 2 to be an exceptionally luminous star with its apparent magnitude due to extreme reddening.

Comparison to the photometric properties of the corresponding population of isolated WN7-9ha and O hypergiants
shows an encouraging co-location in the colour/magnitude plot, although with an increased proportion of fainter, redder examples. This is likely indicative
of greater interstellar reddening along the relevant lines of sight, although verification awaits  a parameterisation of 
the reddening law towards the GC. The newly identified WN8-9ha star S120 appears an exception to this trend, being  the faintest example ($K\sim12.5$) but with a rather moderate near-IR colour ($(H-K)\sim1.6$), suggesting it has an intrinsically low 
luminosity.

\begin{figure*}
\includegraphics[width=14cm,angle=-0]{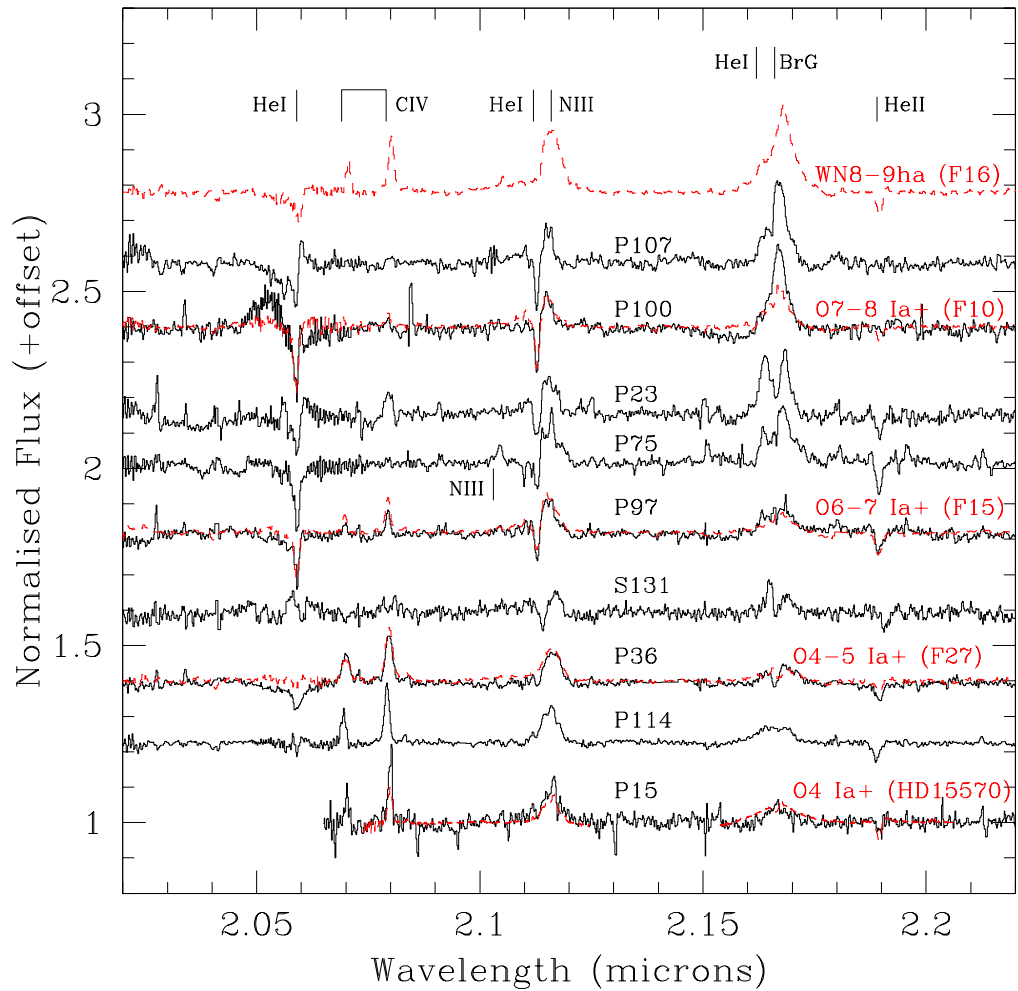}           
\caption{Montage of spectra of isolated O hypergiants. Template spectra  are 
overplotted in red along with identifications; we also show the spectrum of F16, the WN8-9ha star with the weakest emission lines within the Arches (Clark et al. \cite{clark18a}), for comparison to P100 and 107. We note that a narrow  emission feature of uncertain origin at line centre of Br$\gamma$ in the spectrum of P114 was removed, leaving an artificially flat topped profile.}
\end{figure*}

\begin{figure*}
\includegraphics[width=11cm,angle=-90]{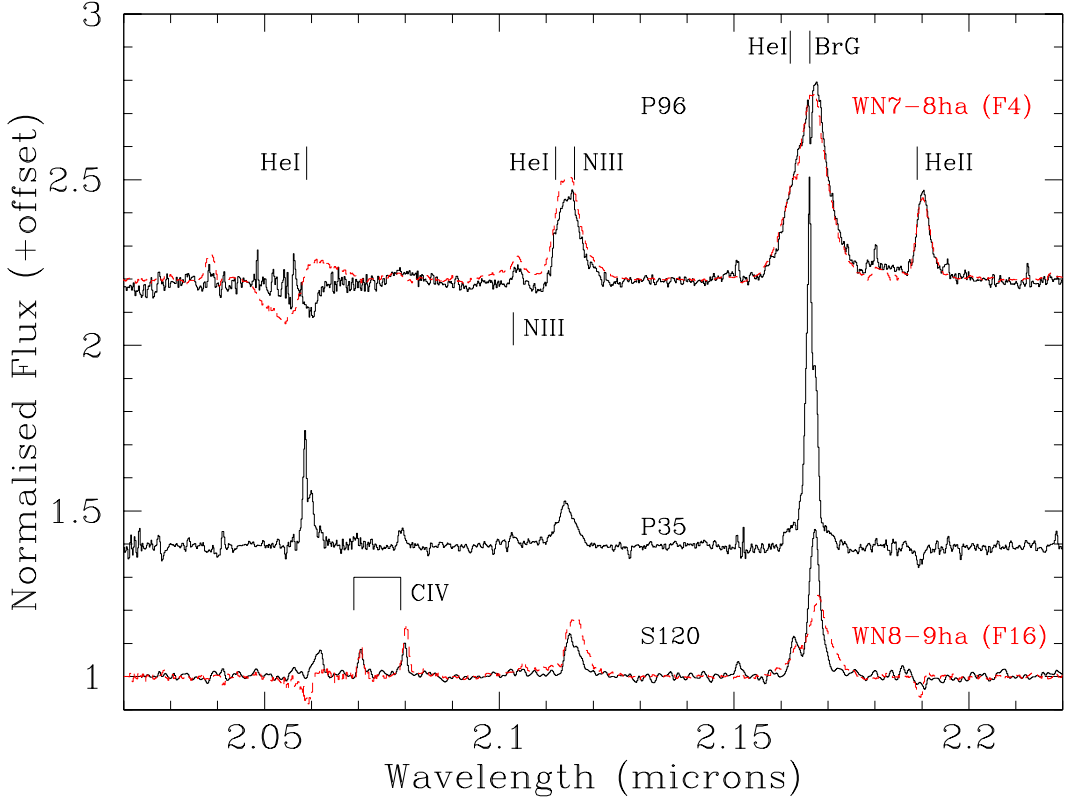}            
\caption{Montage of spectra of WNLha stars. Template spectra with classifications
overplotted in red  (Clark et al. \cite{clark18a}). We note that S120 appears to have an anomalous RV redshift of 
$\sim+100$kms$^{-1}$; the spectrum of F16 has been artificially shifted by a comparable amount to aid in comparison.}  
\end{figure*}

\begin{figure}
\includegraphics[width=9.7cm,angle=-0]{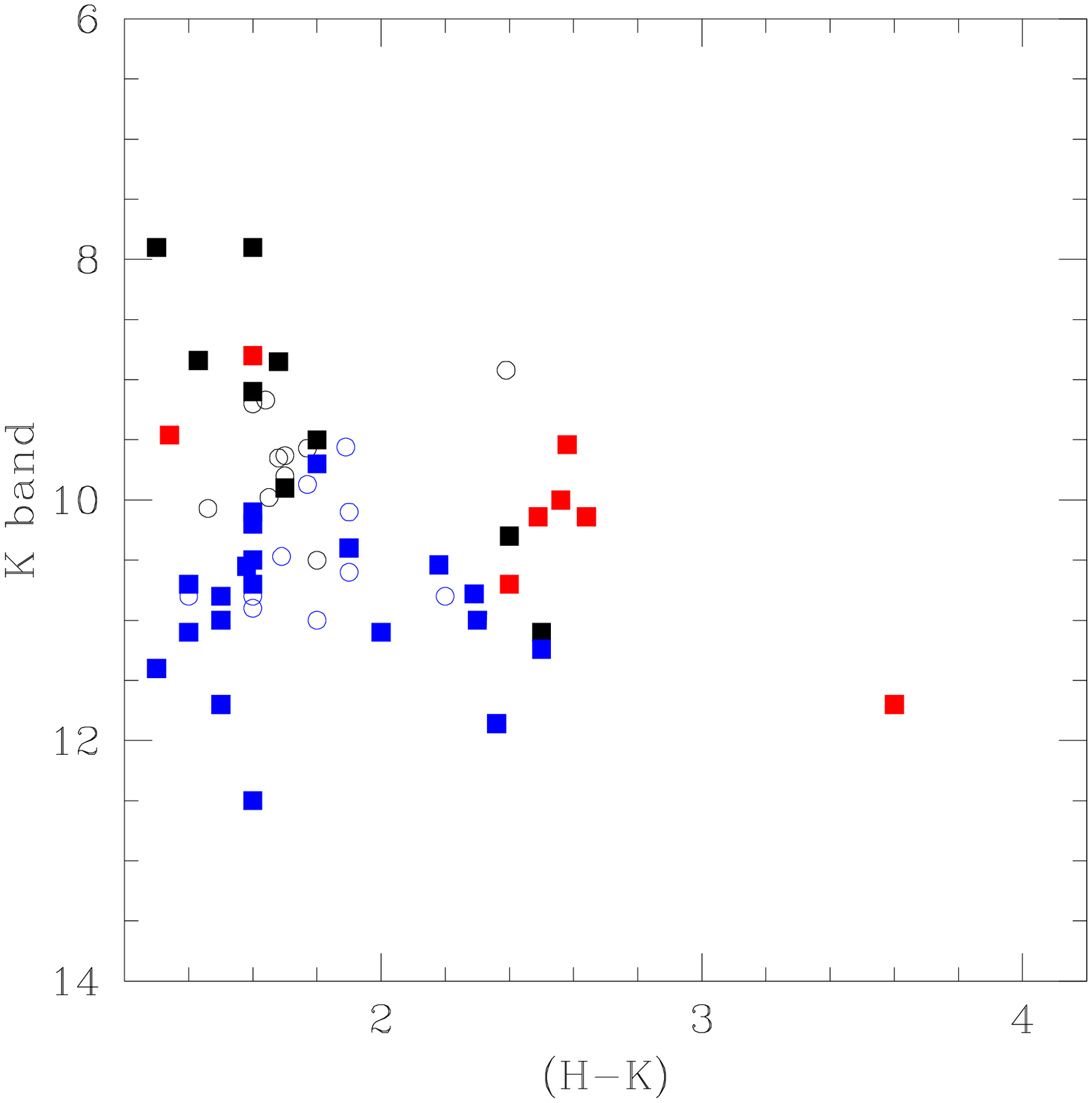}            
\caption{IR colour magnitude diagram for WN7-9ha and mid-O hypergiants (blue symbols), WN9-11h and early-B hypergiants (black symbols) 
and putative cool blue hypergiants and  supergiant B[e] stars (red symbols). 
Relevant members of the Arches (WN7-9ha and mid-O hypergiants) and Quintuplet (WN9-11h and early-B hypergiants)
clusters are given by open blue and black circles respectively, while 
isolated stars are represented by filled squares of the appropriate colour. Axis scale chosen to mirror the comparable plot of 
WN and WC stars (Fig. 10). }
\end{figure}

\subsection{WN9-11h stars/early-B hypergiants}

We present the spectra of WN9-11h stars and early-B hypergiants in Figs. 5 and 6. Given the large number of such stars within 
the Quintuplet cluster compared to those known in the wider galaxy (particularly early-B hypergiants) we utilise these as 
classification templates following the discussion in Clark et al. (\cite{clark18b}). 
In doing so we are able to revise the classification of P103 from generic P Cyg O-type supergiant to 
B0-1 Ia$^+$/WNLh (Fig. 5). Interpreting the spectrum of P56 - plotted against the WN11h star LHO71 in Fig. 6 - is more difficult. Both stars clearly show prominent electron scattering wings in the P Cygni profile of He\,{\sc i}
2.059$\mu$m, while their Br$\gamma$ emission lines are broadly comparable. However the He\,{\sc i} 2.112$\mu$m doublet is  in absorption in P56 but in emission in LHO71; given this discrepancy we suggest that a classification  as either B0-1 Ia$^+$  or WN11h would be appropriate (cf. P103).  Likewise P98 and P137 undergo less dramatic revisions to B1-2 Ia$^+$/WNLh  and WN10h respectively, with P19
found to be a twin of the broad lined WN9h Quintuplet member LHO 158 (cf. Clark et al. \cite{clark19a}). Of these we note that Hankins et al. (\cite{hankins}) report P137 is located  within a mid-IR ring nebula, possible indicative of a wind blown bubble or circumstellar ejecta (cf.
 the Pistol star).

The combination of He\,{\sc i} 2.059$\mu$m 
and Mg\,{\sc ii} 2.138/44$\mu$m emission and narrow Br$\gamma$ and He\,{\sc i} 2.112 and 2.161$\mu$m absorption in the spectrum of 
S132 indicates that it is a new early- B hypergiant (Fig. 5 and Table 1).
Comparison of the spectrum of SSTU J174523.11-290329.3 - which also demonstrates Mg\,{\sc ii} emission (Mauerhan et al. \cite{mauerhan07}; their Fig. 4)  - to those of S132 and similar objects within the Quintuplet suggests that an identification as a  hypergiant, rather than the previous  supergiant classification, is more appropriate. Finally, despite the low S/N and resolution of the spectrum of the previously unclassified 
2MASS J17461292-2849001 (Geballe et al. \cite{geballe}; their Fig. 6), it appears directly comparable to the preceding two stars. As a consequence 
we adopt a similar B1-3 Ia$^+$ classification for it; further strengthened by its close proximity to - and hence potential membership of -  the Quintuplet,  which hosts a large number of such stars (Clark et al. \cite{clark18b}).

Fig. 4 illustrates the near-IR photometric properties of both isolated early-B hypergiants and WN9-11h stars and their counterparts within the Quintuplet cluster. The majority of Quintuplet members occupy a relatively compact region of colour/magnitude space ($9{\lesssim}K{\lesssim}10$ and $1.4{\lesssim}(H-K){\lesssim}1.8$)\footnote{An identical lower bound to the colour $(H-K)$ index is also seen for the Arches cluster. This is substantially in excess of the limit of $(H-K){\gtrsim}1.0$ adopted by Dong et al. (\cite{dong12}) for early type stars within the CMZ. However, since  many of these stars  are expected to show an intrinsic IR excess due to continuum emission from their stellar winds we caution against accepting $(H-K){\gtrsim}1.4$ as a blanket colour cut for all massive stars in the CMZ.}. Outliers include the faint ($K\sim10.5$) WN9h star LHO 158 and the extremely red WN10h outlier 
LHO 67.The former is likely the hottest of this cohort (and hence may require the largest bolometric correction) while an understanding of the latter  - intrinsic IR excess and/or extrinsic reddening - awaits detailed quantitative analysis.  

While five of the  isolated  early-B hypergiants and WN9-11h 
stars\footnote{P56, 98 \& 137,  2MASS J17461292-2849001 and WR102ka} 
are co-located with  Quintuplet members  in the colour/magnitude plot, four are outliers. Both P19 (WN9h) and SSTU J174523.11-290329.3
(B0-2 Ia$^+$) appear rather faint and red and  likely suffer excess interstellar reddening  (cf. Arches F2; Sect 3.2). Conversely the 
bright WN10h star P137 may either be seen through a window suffering reduced extinction  or is a foreground object. Finally despite being over a magnitude brighter than any other cluster or isolated early-B hypergiant observed to date, the $(H-K)$ colour of S132 is 
unexceptional, suggesting that it may be intrinsically highly luminous.

\begin{figure*}
\includegraphics[width=11cm,angle=-90]{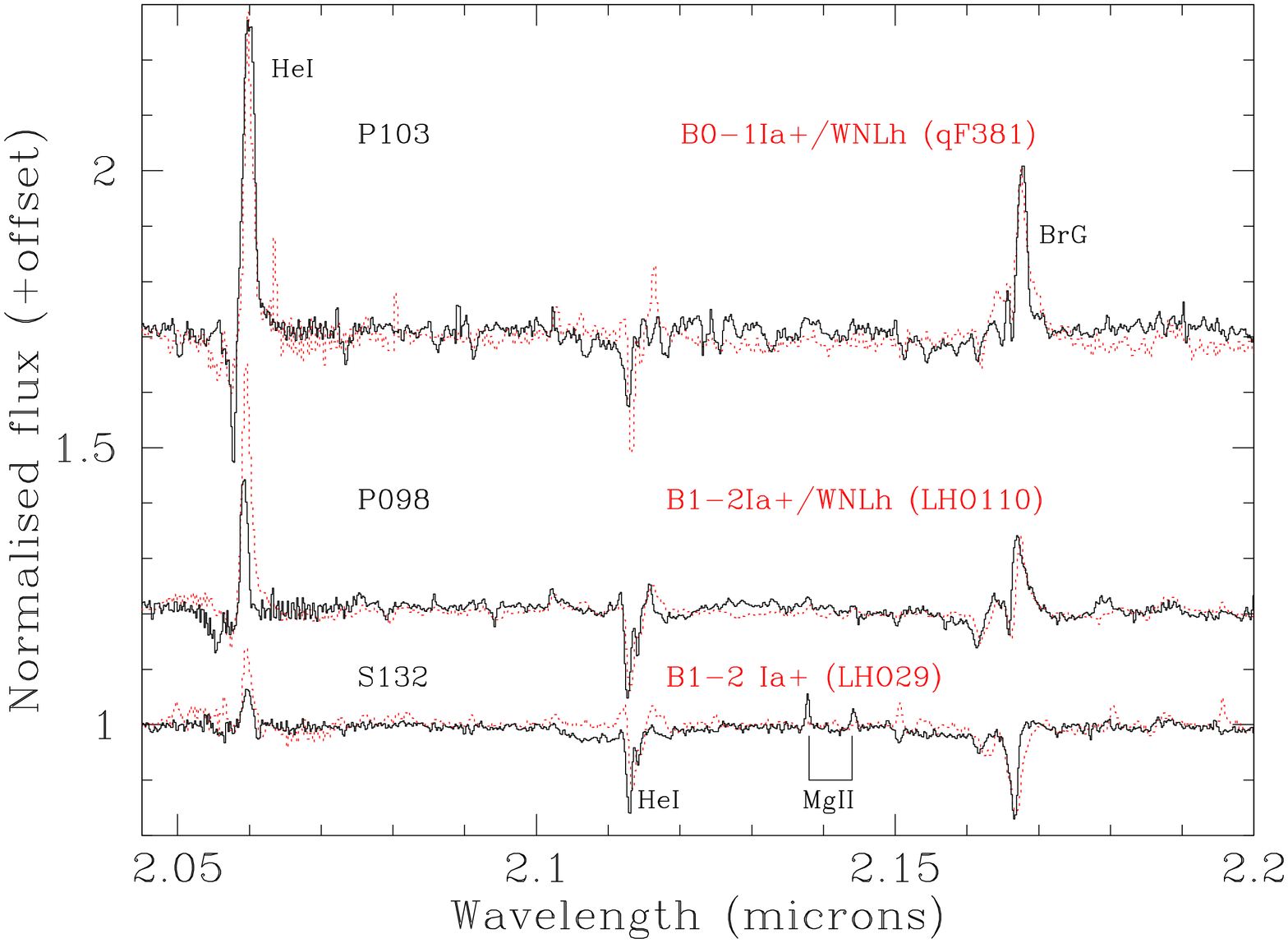}
\caption{Spectra of candidate isolated early-B hypergiants. Comparator spectra of Quintuplet cluster members
and appropriate classifications given in red (Clark et al. \cite{clark18b}).} 
\end{figure*}

\begin{figure*}
\includegraphics[width=11cm,angle=-90]{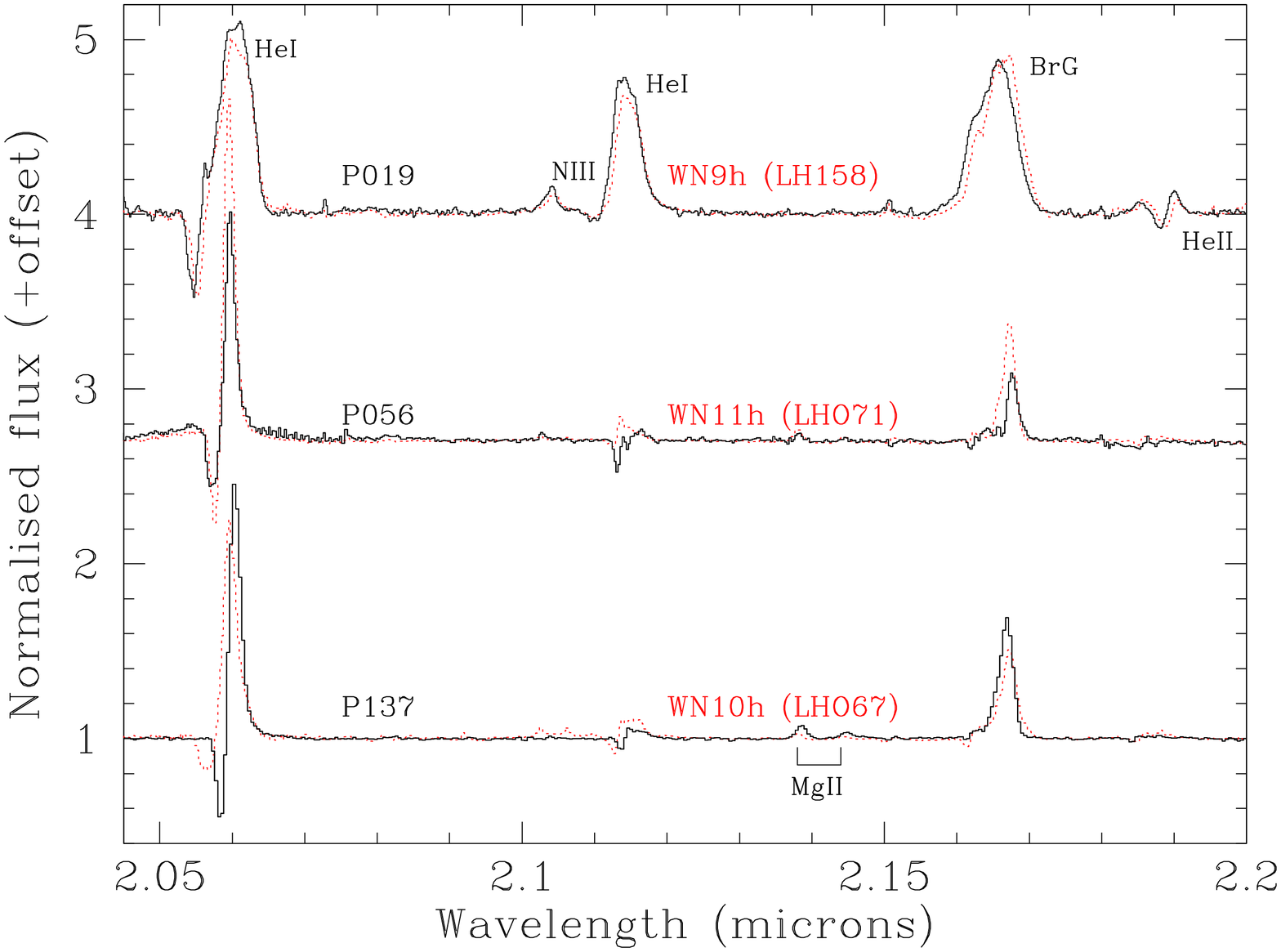}            
\caption{Spectra of WN9-11h stars. Comparator spectra of Quintuplet cluster members
and appropriate classifications given in red (Clark et al. \cite{clark18b}).}
\end{figure*}

\subsection{Candidate cool BHGs/LBVs and  supergiant B[e] stars}

Geballe et al. (\cite{geballe}; their Figs. 5 and 6) report on a cohort of ten stars with spectra dominated by strong Br$\gamma$ and weaker emission in He\,{\sc i} 2.059$\mu$m and various Fe\,{\sc ii} transitions.  Of these 2MASS J17452861-2856049 and J17462830-2839205 
correspond to P35 (WN8-9ha +neb) and CXOGC J174628.2-283920 (O4-6 Ia) respectively (Table 1), while 2MASS J17455154-2900231 is source D of Cotera et al. (\cite{cotera99}; Sect 3.8). This leaves a total of seven objects of which, fortuitously, we have observations of two -  P112 and 141 (=2MASS J17453782-2857161 and J17450929-2908164 respectively) - allowing us to verify line identifications  from spectra of greatly improved S/N and resolution. These data are presented  in Fig. 7, along with the spectrum of a third object of similar morphology -  P40 (=2MASS J17452405-2900589). 

Prior to discussing spectral morphologies it is instructive to consider photometric data. 
 Only two of the stars - P40 and 2MASS J17470940-2849235 - have $(H-K)< 2.0$, with the
latter sufficiently blue that a foreground nature cannot be excluded (Table 1).
Five of the remaining  objects cluster between $(H-K)\sim2.4-2.6$ with the sixth, P141, an outlier with 
$(H-K)\sim3.6$; values significantly in excess of those of other early-type massive stars in the GC (Table 1 and Fig. 4). 
Pre-empting the following discussion, in the absence of a classification yielding `photospheric' colours for this cohort it is impossible to quantitatively decouple intrinsic and extrinsic contributions to reddening, although it seems highly likely that a significant continuum  contribution from a  wind or dusty circumstellar disc is present in these six stars. Such a conclusion is supported by mid-IR observations,
with a number of stars\footnote{P40, 112, \& 141, 2MASS J17444319-2937526 and J17470940-2849235.} appearing intrinsically red upon consideration of  the $[4.5]-[8.0]>1.0$ colour cut suggested by Robitaille et al. (\cite{rob}; we refer the reader to this work for a detailed justification of this criterion). Unfortunately, the parallel possibility of  significant interstellar extinction affecting the near-IR photometry  precludes us from employing the colour-magnitude and colour-colour plots of, for example,  
 Bonanos et al. (\cite{bonanos}) in order to determine the physical nature of these stars. 

Mindful of these issues we now turn to the spectra of P40, 112 and 141. All three are dominated by strong, narrow Br$\gamma$ emission. Critically, broad electron scattering wings are also seen in the profile of this transition in P40 and 112; indicative of a dense stellar wind and hence a massive star identification. Weak He\,{\sc i} 2.059$\mu$m emission is also apparent in the spectra of both these stars, suggesting they are hotter than P141 (where it is absent), although no trace of the He\,{\sc i} 2.112$\mu$m feature is present in any of the three, nor are the spectral signatures of high excitation species  such as He\,{\sc ii}, N\,{\sc iii} or C\,{\sc iv}. Instead the remaining emission features present arise from low excitation metal transitions such as Fe\,{\sc ii} 2.061$\mu$m (P40 \& 112)  and  2.089$\mu$m (all stars), 
[Fe\,{\sc ii}] 2.045$\mu$m, 2.118$\mu$m, and 2.133$\mu$m (P141), Mg\,{\sc i} 2.134+2.144$\mu$m (P112) and Na\,{\sc i} 2.206+2.209$\mu$m (P40). At longer wavelengths the CO bandheads are seen in emission in P40 (Fig. 8); they and the Pfund series are absent from both P112 and P141. Finally,  there is no indication of H$_2$ emission in any of the stars, disfavouring a pre-MS classification. 

The narrow emission line spectra dominated by Br$\gamma$ and low excitation metals are reminiscent of both cool-phase LBVs
(Clark et al. \cite{clark11}, \cite{clark18b}) and supergiant B[e] stars (sgB[e]; Oksala et al. \cite{oksala}). CO bandhead emission is present in a substantial number of sgB[e] stars but appears absent from most, if not all, LBVs (Morris et al. \cite{morris}, Oksala et al. \cite{oksala}). We present  the spectra of the LBVs FMM362 and G24.73+0.69 in Fig. 7 to illustrate the gross similarities to P40, 121 and 141, although neither star provides an exact match. As a consequence
we suggest a sgB[e] classification for P40 and, pending an evaluation of long term variability, either a late-B hypergiant or cool LBV classification for P112 and 141, under the assumption they are located within the CMZ; estimation of stellar luminosities and temperatures will have to await quantitative model atmosphere analysis. 

Given that 2MASS J17444319-2937526, J17445538-2941284 and 2MASS J17450241-2854392 are co-located with P112 in the near-IR colour/magnitude plot (Fig. 4) and appear to show a comparable spectral morphology (subject to the low resolution and S/N data) it is tempting to apply a similar classification to them. While the same is true for 2MASS J17482472-2824313,  we are more cautious in this case due to a possible association with a cold, dusty clump (Contreras et al. \cite{contreras}) which could favour a pre-MS status. Likewise the comparatively  blue $(H-K)$ 
colour for 2MASS J17470940-2849235  leaves open the possibility of a lower mass foreground (post-AGB) object. Indeed, it is entirely possible that this cohort could be rather heterogeneous -   comprising stars of different luminosities and evolutionary status but similar gross observational features. This would be analogous to stars exhibiting the B[e] phenomenon (cf. Lamers et al. \cite{lamers}) which, as demonstrated by P40, these stars closely resemble. 

Finally we note that based on its spectral morphology the LBV G0.120-0.048  would be included  in 
this cohort had not its proximity to the Quintuplet cluster ($\sim7$pc distant) suggested possible cluster membership (Mauerhan et al. \cite{mauerhan10b}; footnote 11).

\begin{figure*}
\includegraphics[width=11cm,angle=-90]{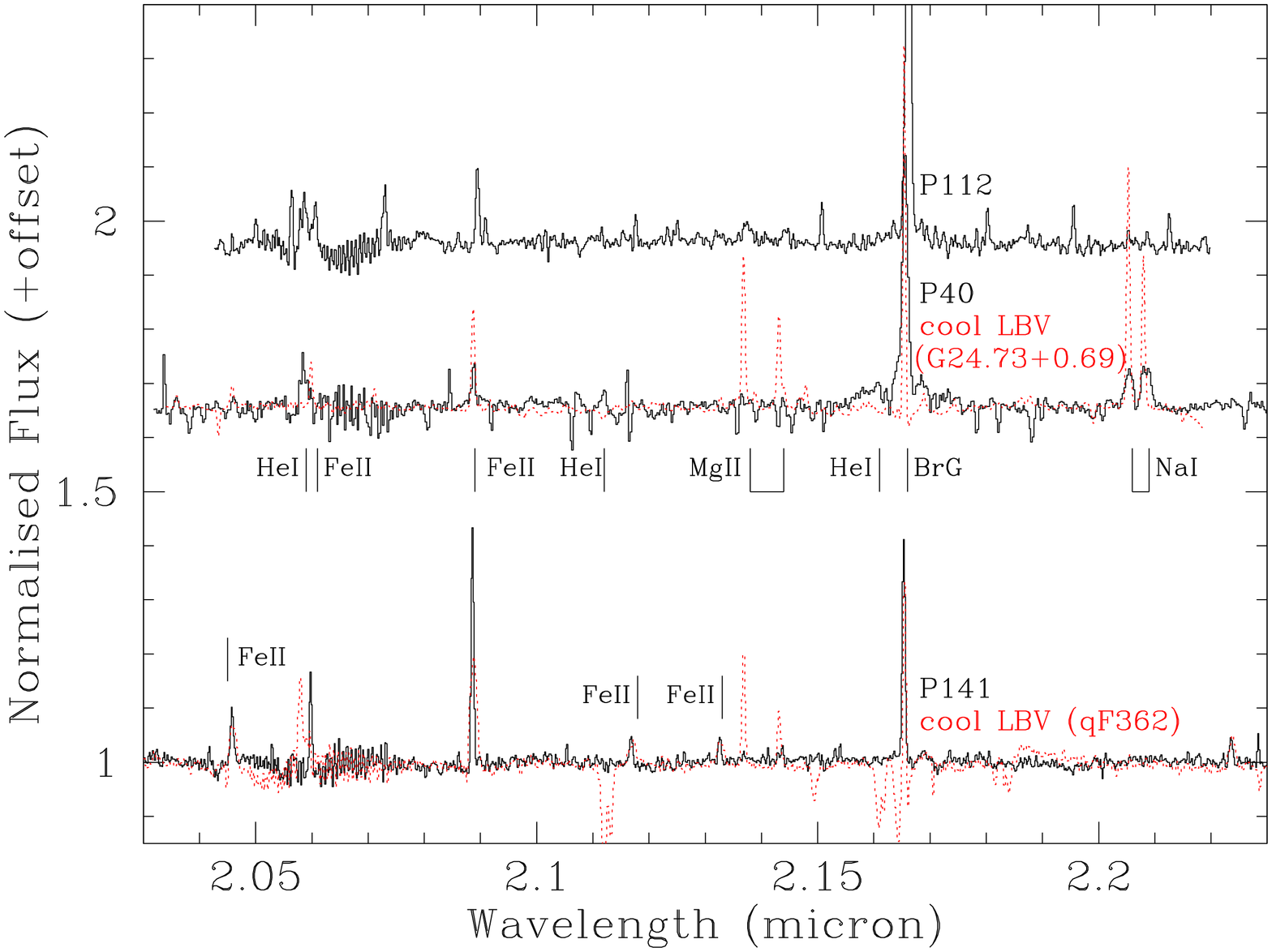}            
\caption{Montage of spectra of isolated candidate late-B hypergiants/cool LBVs and sgB[e] stars.
Spectra of the LBVs qF362 and G24.73+0.69 (Clark et al. \cite{clark18b}) shown for comparison.}
\end{figure*}

\begin{figure}
\includegraphics[width=7cm,angle=-90]{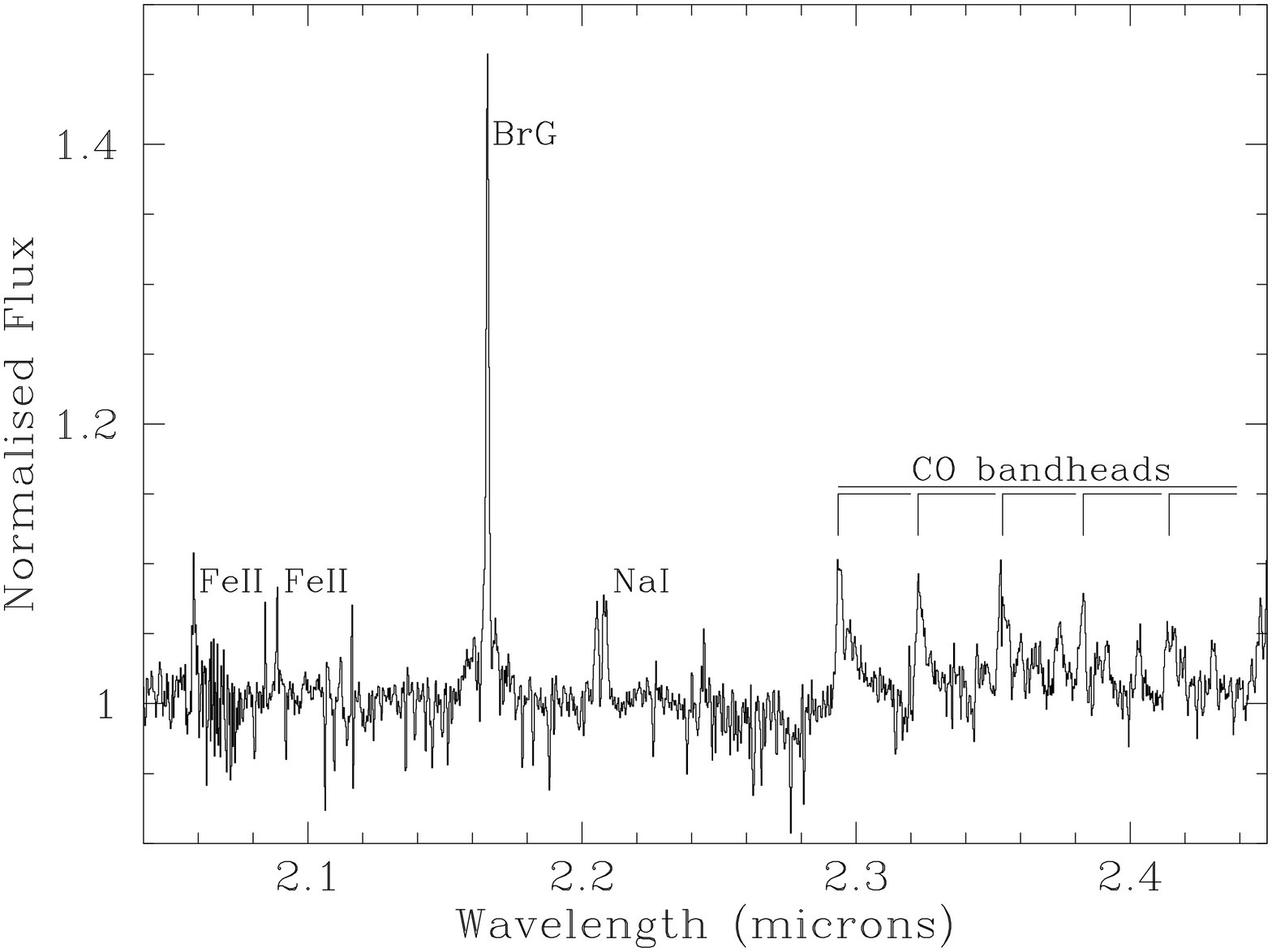}            
\caption{K-band spectrum of the sgB[e] star P40 plotted with an increased wavelength coverage to 
illustrate the presence of CO bandhead emission.}
\end{figure}

\subsection{WN5-7 stars}

Eight isolated WN5-7 stars have been identified within the GC, of which  two - P99 and P150 -  are new discoveries (Table 1). A ninth - qF353 (=P64; Steinke et al. \cite{steinke}) - is located on the periphery of the apparent wind blown structure encircling the north and east quadrants of the  Quintuplet cluster; since it has historically been associated with this cluster, we do not include it in this census. We note that  P2 (=[MCD2010] 17) - one of the  eight stars considered here - is also proximate to the Quintuplet (Mauerhan et al. \cite{mauerhan10a}).

The spectra of the seven stars presented in Fig. 9 are all dominated by strong and broad emission in He\,{\sc i}, He\,{\sc ii} and N\,{\sc iii}. The  line widths of P39 and 91 support a classification as broad lined systems; the low resolution spectrum of P2 suggesting likewise (Mauerhan et al. \cite{mauerhan10a}; their Fig. 4).

Steinke et al. (\cite{steinke}) identify qF353E as a WN6 star; informed by Rosslowe \& Crowther (\cite{rosslowe}) we use this as a benchmark to provide relative classifications for the remaining stars. The close similarity of P34 to qF353E  suggests an identical classification for this star. In comparison to qF353E, the He\,{\sc ii} 2.189$\mu$m line in P2, 39, 91, and P99 is stronger relative to the other emission features in their spectra, suggesting a WN5-6 identification for these stars (with the lack of N\,{\sc v} 2.10$\mu$m precluding earlier sub-types); conversely the relative  weakness of this transition in P109 and 147 implies a WN6-7 classification. Finally the combination of exceptionally  strong He\,{\sc ii} 2.189$\mu$m emission in P150 and the absence of the  absorption component due to He\,{\sc i} 2.059$\mu$m that is present in all other examples suggest that this is the hottest star observed and hence we assign a WN5 sub-type.

We present photometry for these stars in Fig. 10, noting that the newly discovered P99 and P150 are the faintest examples identified to date.  Given their near-IR colours P150 appears intrinsically rather faint while P99 could suffer considerable excess extinction along its line of sight. The broad lined WN5-6 star P39 is also noteworthy in this regard, with the most extreme value of the cohort ($H-K \sim3$), although it does not appear exceptionally faint ($K\sim12.2$); if the near-IR colour is due to substantial  interstellar reddening it would appear intrinsically highly luminous.

\begin{figure*}
\includegraphics[width=14cm,angle=-0]{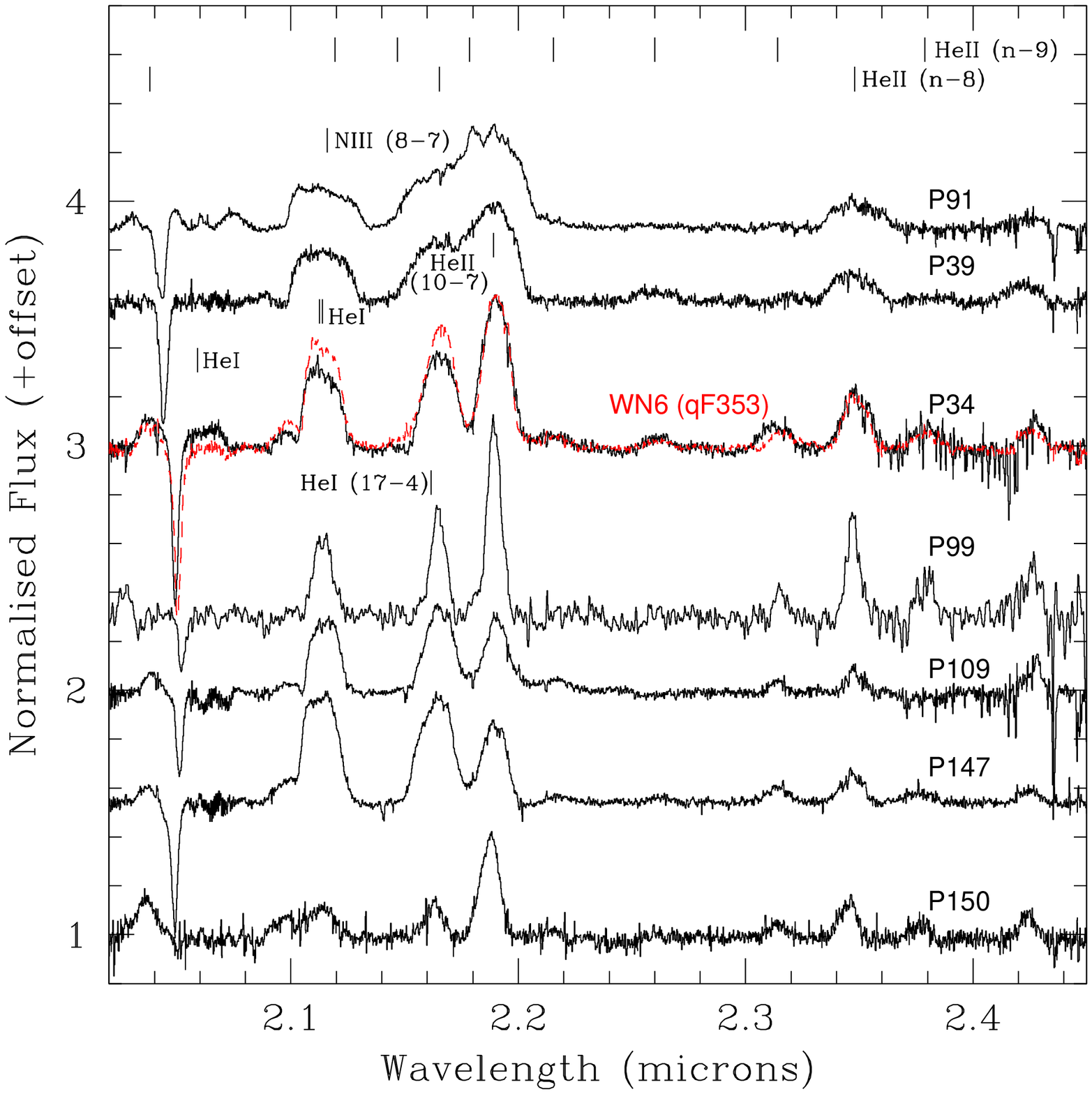}            
\caption{Montage of spectra of isolated candidate WN5-7 Wolf-Rayets. The spectrum of the WN6 star qF353E is shown in red for comparison.
Prominent transitions are indicated; for completeness the  unlabelled He\,{\sc i} lines at 2.059$\mu$m and 2.1126+2.1138$\mu$m are the 2p$^1$P$^{\rm o}$ - 2s$^1$S and the 4s$^3$S - 3p$^3$P$^{\rm o}$ + 4s$^1$S - 3p$^1$P$^{\rm o}$ transitions respectively.}
\end{figure*}

\subsection{WC8-9 stars}

Inspection of the relevant literature (e.g. Mauerhan et al. \cite{mauerhan10a}, \cite{mauerhan10c}, and associated errata;  
Geballe et al. \cite{geballe}) reveals  
13 isolated stars classified as WC stars\footnote{We note that the proximity of WR102ca and CXOGC J174617.7-285007 to the Quintuplet cluster suggest a physical association; hence we do not consider these members of the isolated WC star cohort.}. As with the Quintuplet cohort (Clark et al. \cite{clark18b})\footnote{The inclusion of CXOGC J174617.7-285007 brings the cluster total at the time of writing  to 14.} these are all WC8 and WC9 stars; indeed only a single example of an earlier subtype - the WC5/6 star IRS 3E (associated with Galactic Centre cluster; Paumard et al. \cite{paumard}) - has been identified within the CMZ. We obtained spectra of six examples which are presented in Fig. 11. Despite improved S/N and resolution, our spectra of P28, 49, 53, 101, and 151 
do not necessitate re-classification, although they do illustrate two essential features of this cohort:  that stars of  the same sub-type present spectra with unexpectedly diverse morphologies - presumably indicative of differences in both stellar and wind properties - and that dilution/obscuration  of emission features by a substantial near-IR continuum excess due to the presence of hot dust is common. 

The spectrum of the final object, MP13 (=CXOGC J174519.1-290321; WC9d), exemplifies the latter phenomenon, with the  weak emission line spectrum  indicating  the presence of hot dust. Intriguingly, the He\,{\sc ii} lines  that are evident are suggestive of a  WN7 classification (Fig. 9), with none of the weaker C\,{\sc ii-iv} features that characterise WCL stars being visible, despite their presence in the spectrum of Mauerhan et al. (\cite{mauerhan10c}; their Fig. 6). While we retain a  WC9d classification for this object on the basis of the latter work, we suspect that it may be intrinsically variable, with greater dilution in our spectrum compared to that of Mauerhan et al. (\cite{mauerhan10c}) - although we cannot exclude the possibility of a WN7 companion at this time.

Within the GC, excess continuum emission from hot dust was first recognised for five exceptionally bright near-IR sources within the 
Quintuplet cluster. High spatial resolution near-IR imaging revealed these to be colliding wind binaries 
(Tuthill et al. \cite{tuthill}), with subsequent high S/N and resolution $JHK$ spectroscopic observations identifying   strongly diluted emission  features characteristic of WCL stars (Najarro et al. \cite{paco17}). 

Despite its featureless $K-$band spectrum  Mauerhan et al. 
(\cite{mauerhan10c}) assigned  a WCLd classification to CXOGC J174645.2-281547 by analogy to the Quintuplet cohort and by virtue of its hard X-ray emission. Geballe et al. (\cite{geballe}; their Fig. 2) report five further objects with very red, essentially featureless $K-$band spectra; however, despite the possible presence of weak He\,{\sc i} 2.059$\mu$m emission in two examples they refrain from classifying these  as dusty WCL  stars.

We may question whether it is  possible to provide a more  definitive classification of CXOGC J174645.2-281547 and the five sources from Geballe et al. (\cite{geballe}) via consideration of photometry and other observational data. Of these 2MASS J17431001-2951460 appears associated with the methanol maser MMB G358.931-0.030 (0.3" distant; Caswell et al.
\cite{caswell77}, \cite{caswell10}) which argues for an object in a pre-MS  evolutionary phase. 2MASS J17445461-2852042 (= MGM 1 1, [MKN2009] 7) is a large-amplitude photometric variable with an unconstrained, but apparently long period (${\Delta}H\sim0.64$,
${\Delta}K\sim1.12$; Monte et al. \cite{mon92}, Matsunaga et al. \cite{matsunaga}); as such we exclude it from the colour/magnitude plot noting that it is most likely a Mira variable. This leaves three remaining candidate  WCLd stars - 2MASS J17432173-2951430, J17432988-2950074, and 2MASS J17460215-2857235 - which we plot along with the remaining isolated WCL stars and, for context,  the Quintuplet cohort in Fig. 10.

Comparison to other evolutionary groupings reveals that the  WCLd stars are the most photometrically diverse, with $K\sim6.5-12.7$ 
(Figs. 4 \& 10). There is a significant scatter in $(H-K)$ colour index but an overall correlation - in the sense that brighter sources are redder - is present. While multiple physical causes are clearly implied - such as intrinsic differences in stellar and wind properties and  differential interstellar reddening - we suppose that an increasing  contribution from hot circumstellar dust dominates this relationship; a conclusion supported by the fact that the brightest and reddest sources within the Quintuplet cluster are those with essentially  featureless spectra due to dust dilution.

It is therefore encouraging that the spectroscopically confirmed, isolated WCLd stars plotted in the colour/magnitude diagram are coincident with examples found within the Quintuplet, while  CXOGC J174645.2-281547, 2MASS J17432173-2951430, J17432988-2950074, and 2MASS J17460215-2857235
seamlessly extend this co-location to brighter, redder objects with featureless spectra\footnote{The final star with a featureless spectrum from  Geballe et al. (\cite{geballe}) - 2MASS J17431001-2951460 - is $\sim2.5$ magnitudes fainter and has a significantly bluer $(H-K)$ colour index than the preceding objects, bolstering the conclusion that it
is in a pre-MS evolutionary phase  given its association with a methanol maser.}. Moving to fainter $K-$band magnitudes and, of the three isolated WC stars with $K\sim9-9.5$, the emission lines in the spectra  of the two stars with the largest
$(H-K)$ values  - 2MASS J17444083-2926550 and J17463219-2844546 -  are also very weak (Geballe et al. \cite{geballe}; their Fig. 4). This is consistent with the presence of substantial continuum veiling due to emission from hot dust, an hypothesis strengthened by their extremely red $(J-K)$ colour indices. 

Given this we conclude that, as assumed for CXOGC J174645.2-281547, the three isolated objects from Geballe et al. (\cite{geballe}) with featureless $K-$band spectra  are also bona fide dusty WCL stars. Indeed we may invert the argument: given that the Quintuplet clearly hosts such stars one would anticipate their presence in the isolated stellar cohort, so if these objects are not dusty WCL stars one would be need to explain their absence. 
In either eventuality we close this discussion by noting that the number of isolated dusty WCs within the CMZ - 13 or 16, depending on the nature of these sources -  is directly comparable to the number associated with the Quintuplet cluster (see footnote five).

\begin{figure}
\includegraphics[width=9.7cm,angle=-0]{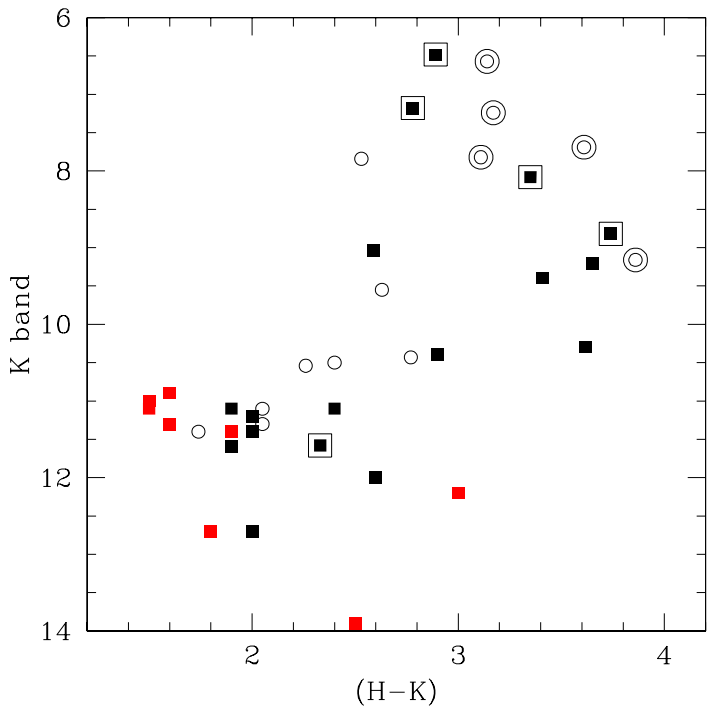}            
\caption{IR colour magnitude diagram for candidate and confirmed WN5-7 (red) and WCL stars (black).
Members of the Quintuplet cluster and isolated examples given by open circles and filled squares respectively. 
Stars with featureless K-band spectra  are 
indicated by nested symbols, the anomalously faint example being 2MASS J17431001-2951460 (with  2MASS J17445461-2852042 excluded due to variability; Sect. 3.8).  
Ground-based photometry for the Quintuplet members given in Cutri et al. (\cite{cutri}), Dong et al. (\cite{dong12}), 
and Hussmann et al. (\cite{hussmann}). No photometry is available for qF235N while the location of qF76 \& 309 
($K\sim11.2, (H-K)\sim2.0$ and $K\sim13.4, (H-K)\sim2.0$ respectively) have been displaced slightly to avoid overlap with
P28 and 101.}
\end{figure}

\begin{figure*}
\includegraphics[width=14cm,angle=-0]{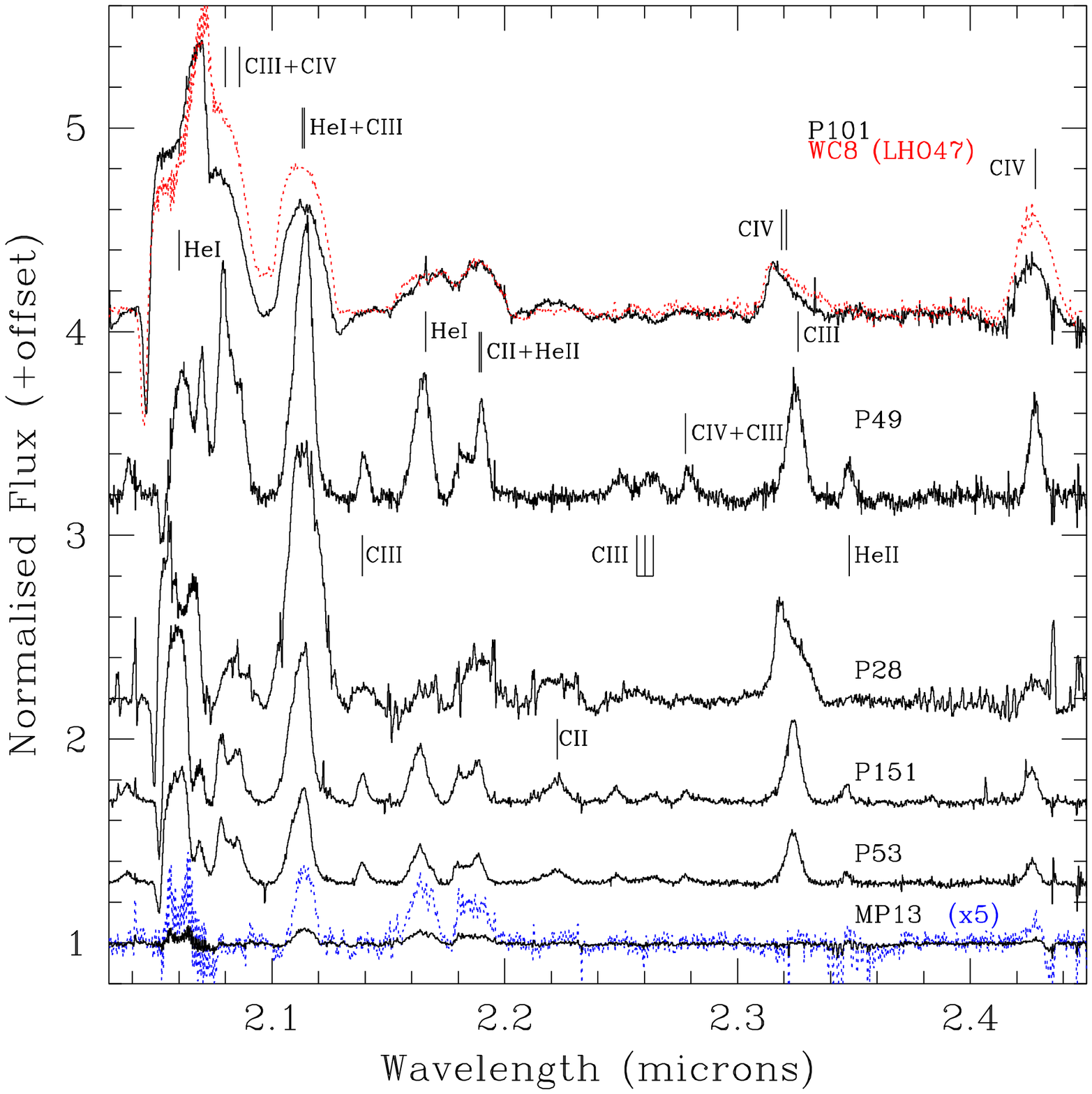}            
\caption{Montage of spectra of isolated WC8-9 Wolf Rayet stars, with the Quintuplet WC8 star LHO 47 plotted in red for 
comparison (Clark et al. \cite{clark18b}). Given the remarkable diversity of morphologies, no additional  WC9 comparator spectra from the
Quintuplet are available.  The emission features in MP13 are particularly weak; we therefore enhance these by a factor of 
five and overplot the resultant spectrum in blue for clarity.}
\end{figure*}

\subsection{Classical Oe/Be stars}

The last homogeneous cohort  that may be identified in our observations are a group of six faint ($K\sim13-14.6$) stars  
with spectra dominated by strong Br$\gamma$ emission and, in a subset, weak He\,{\sc i} 2.059$\mu$m emission (Fig. 12).
He\,{\sc i} and He\,{\sc ii} photospheric lines - which might enable a temperature determination - are absent, as is emission in both high and low excitation metallic transitions (although the Mg\,{\sc ii} 2.138+2.144$\mu$m doublet was identified in the
 spectrum of P105 presented by de Witt et al. \cite{dewitt}). The lack of low excitation atomic (Na\,{\sc i}, Ca\,{\sc i}) or molecular (CO bandhead) absorption features disfavours  the possibility that these are foreground cataclysmic variables. Moreover, the absence of CO bandhead emission distinguishes these stars from the IR excess objects within the Arches cluster, which Stolte et al. (\cite{stolte10}) suggest are B-type stars surrounded by remnant protostellar discs. Instead they most closely resembles classical Oe/Be stars (Clark \& Steele \cite{clark00}); rapidly rotating non-supergiant late-O to early-A stars, characterised by  gaseous, quasi-Keplerian
circumstellar decretion discs that generate line emission in H\,{\sc i}, He\,{\sc i} and low excitation metallic transitions as well as a 
near-IR continuum excess (cf. Porter \& Rivinius \cite{PR}). 

We may ask whether the observational properties of this cohort are consonant with such a classification. The equivalent widths, full-width half maxima and line intensities of the Br$\gamma$ and - where present - He\,{\sc i} 2.059$\mu$m emission lines are consistent with the range expected for early-B stars exhibiting the Be phenomenon (Clark \& Steele \cite{clark00}). Likewise their broad, asymmetric emission profiles are a natural consequence of the  one-armed density waves that commonly perturb the quasi-Keplerian discs of Be stars, with  the double peaked He\,{\sc i} 2.059$\mu$m line profile of S124 clearly indicative of rotation.
Finally such discs are transient phenomena, leading to significant spectral variability on the timescale of years; potentially explaining the disappearance of Mg\,{\sc ii} emission in the four years between the two observations of P105.

Turning to photometric properties and assuming a distance to the GC of $\sim8$kpc and a representative interstellar 
extinction of $A_K \sim3$ one would expect an O9.5V (B3V) star located there to have $K\sim13.4$ ($K\sim15.7$). Furthermore
adopting an indicative  continuum excess of $K{\gtrsim}1$mag due to  emission  from the circumstellar disc 
(e.g. Dougherty et al. \cite{dougherty}) implies that the range of $K-$band magnitudes anticipated for Be stars within the CMZ 
is  consonant with the stellar cohort  considered here. Likewise, comparison of the intrinsic near-IR colours of Be stars 
 ($(H-K)\sim0.0-0.5$; Lada \& Adams \cite{lada}, Dougherty et al. \cite{dougherty}) to the values  exhibited by 
this cohort ($(H-K)\sim1.5-2.0$; Table 1) suggests a degree of interstellar extinction that is fully consistent with 
that expected for sightlines towards the  GC.

We may also essentially invert this argument. Between $\sim10-20$\% of B0-3  stars exhibit the Be phenomenon at Galactic metallicities (Wisniewski \& Bjorkman \cite{wis}) and, as demonstrated above, their properties (an IR excess and strong line emission) favour their detection via surveys such as that of Dong et al. (\cite{dong11}). Consequently it would be surprising if none were to be found in the GC - one would be forced to invoke a rather contrived star formation history that limited the formation of stars of 
$\sim$10-20M$_{\odot}$ over the past $\sim$25Myr (Wisniewski \& Bjorkman \cite{wis}) or assume that physical conditions there preclude the formation of classical Be stars. 

Therefore, even though we may not exclude alternative classifications  - such as pre-MS stars/massive young stellar objects (YSOs; cf. Bik et al. \cite{bik05}, \cite{bik06}) - at this time, we consider an identification of this cohort  as classical Be stars located within the CMZ to be the most compelling explanation for their spectroscopic and photometric properties as reported here.

\begin{figure*}
\includegraphics[width=11cm,angle=-90]{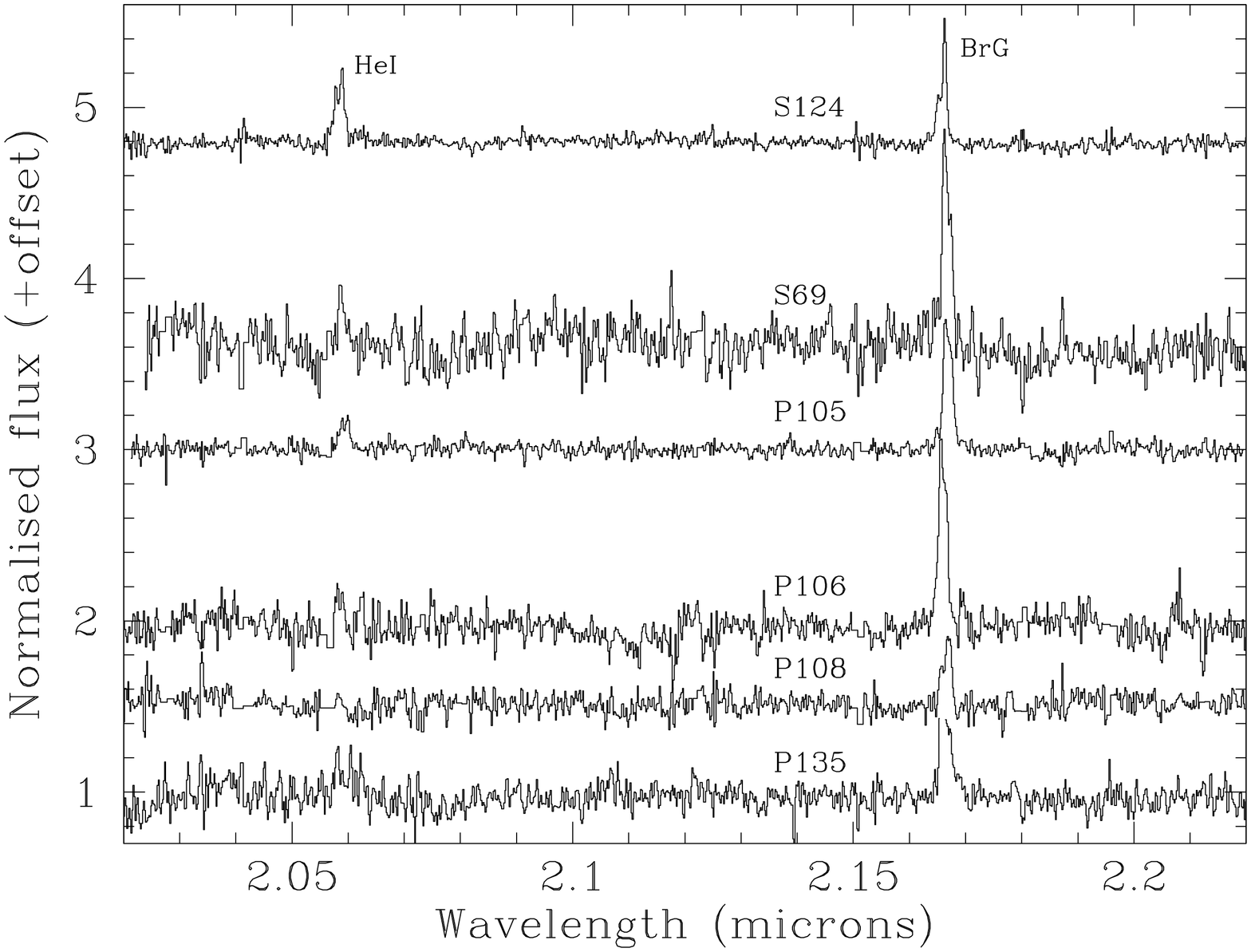}
\caption{Montage of spectra of possible isolated classical Be stars. We caution that the central region of 
the Br$\gamma$ profile of P135 has been artificially removed due to the presence of hot pixels.}
\end{figure*}

\subsection{Miscellaneous and uncertain classification}

\begin{figure*}
\includegraphics[width=11cm,angle=-90]{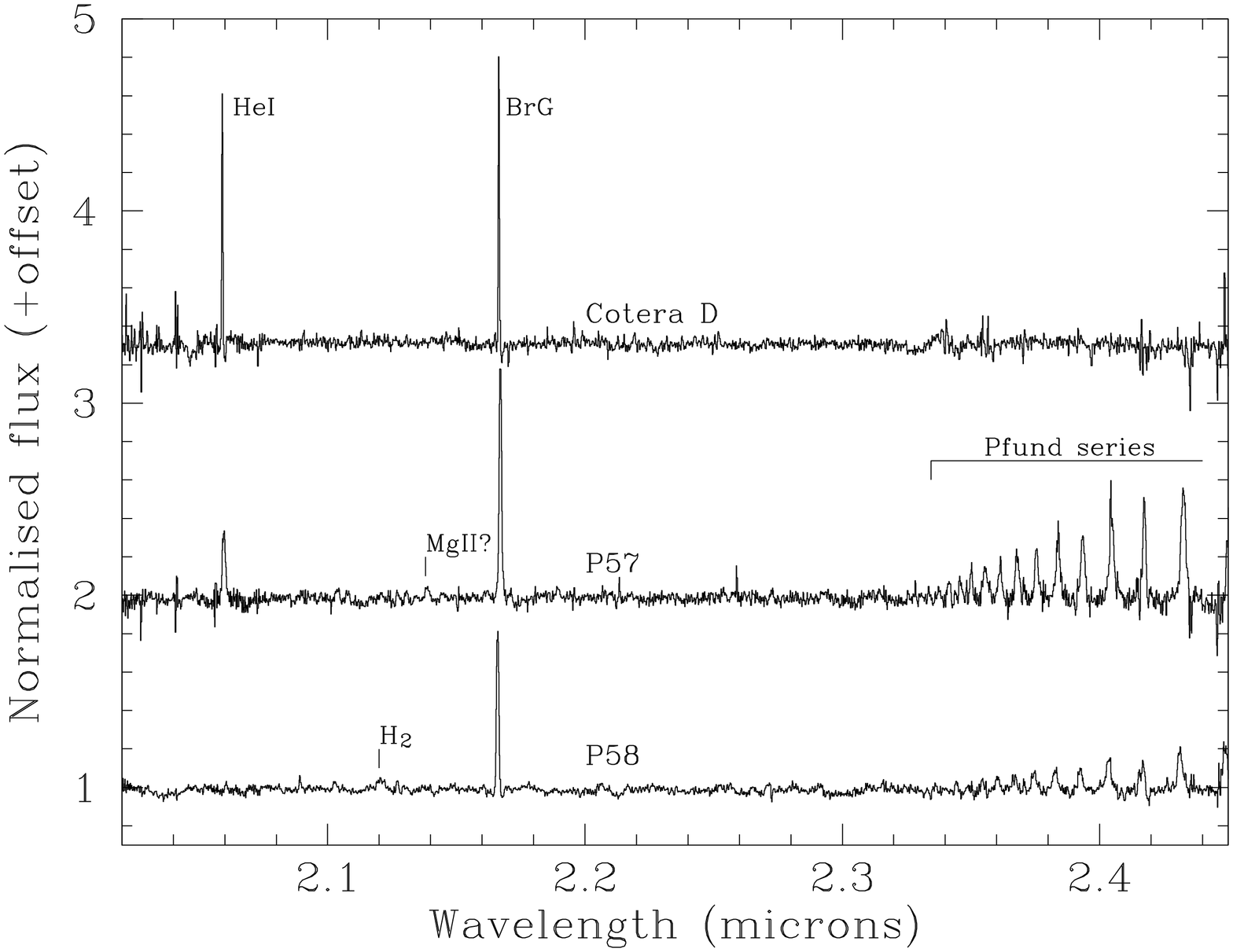}
\caption{Montage of spectra of three sources dominated be narrow H\,{\sc i} recombination lines (Sect. 3.8); the star embedded in the UCH\,{\sc ii} region Sgr A-D, the MYSO/sgB[e] star P57 and  MYSO P58.}
\end{figure*}

Six sources do not fit into any of the above classifications, of which we have new observations of three.
Cotera et al. (\cite{cotera99}) reported that the  spectrum of the highly reddened ($(H-K)\sim3.45$) star associated with the ultra-compact (UC) H\,{\sc ii} region  Sgr A-D was dominated by strong  He\,{\sc i} 2.059$\mu$m and Br$\gamma$ emission lines; 
on this basis they suggested that this indicated a nebular, rather than stellar origin. Despite the increased S/N and resolution of our spectrum (Fig. 13), these remained the only identifiable features. With FWHM$\lesssim100$kms$^{-1}$, the line profiles are narrower than  expected for an origin in either the spherical wind of a hot star or the disc of a classical Be star\footnote{The Br$\gamma$ lines of the six Be stars identified in Sect. 3.7 have FHWM$\gtrsim200$kms$^{-1}$.}; 
we therefore concur with  Cotera et al. (\cite{cotera99}) that the spectrum is dominated by emission from the UCH\,{\sc ii} region.

Both P57 and P58 show strong, single peaked  emission in  Br$\gamma$ and the Pfund series up to at least Pf-30, with FWHM$\sim200$kms$^{-1}$ in our new observations (Fig. 13). Emission 
in the Pfund series is observed in massive YSOs 
(Bik et al. \cite{bik05}, \cite{bik06}), sgB[e] stars (Oksala et al. \cite{oksala}, Kraus et al. \cite{kraus}) and 
LBVs (Oksala et al. \cite{oksala}, Najarro et al. \cite{paco15}). The line widths of both stars are consistent with  any of these possibilities; however the ratio of (Br$\gamma$/Pf-20)$\sim2.5$ for P57 and P58 is significantly lower than that found for for  the Pistol star, where the emission arises in a spherical stellar wind ((Br$\gamma$/Pf-20)$\sim10$; Najarro et al. \cite{paco15}). Hence we conclude that neither star is likely to be an LBV and instead the emission  arises in a circumstellar disc. The presence of weak    
H$_2$ 2.12$\mu$m emission in P58 marks it out as a massive YSO; with $(H-K)\sim1.0$ it is on the cusp of the colour cut applied by Dong et al. (\cite{dong12}) for foreground objects\footnote{Since   pre-MS objects are typically intrinsically red (cf. 2MASS J17431001-2951460 and J17470921-2846161; Table 1) it is possible that P58 is an interloper.}. No H$_2$ emission is seen in P57, while the marginal detection of Mg\,{\sc ii} 2.138$\mu$m is consistent with either a massive YSO or a sgB[e] star; as a consequence we leave both options open at this time. However, in either eventuality the presence of pronounced He\,{\sc i} 2.059$\mu$m emission implies a rather hot source in order to provide the requisite UV photons to drive the line into emission.

Of the remainder 2MASS J17431001-2951460 has already been briefly discussed in Sect. 3.6 where, despite the featureless spectrum presented by 
Geballe et al. (\cite{geballe}; their Fig. 2), near-IR photometry is discrepant with expectations for a dusty WCL star; rather the presence of the methanol maser source MMB G358.931-0.030 only 0.3" away (Caswell et al. \cite{caswell77}, \cite{caswell10}) is suggestive of a YSO classification. 
The presence of CO bandhead emission associated with 2MASS J17470921-2846161 (Geballe et al. \cite{geballe}; their Fig. 6) implies a dense, cool circumstellar envelope; the addition  of H$_2$ molecular emission further points towards a (M)YSO classification (Bik et al. \cite{bik05}, \cite{bik06}). Unfortunately, given the potential for a near-IR continuum excess from the circumstellar envelope and uncertain interstellar extinction it is not possible to make any further inferences as to the nature of the source from extant photometry at this time. 

Finally  2MASS J17444840-2902163 has a spectrum dominated by Br$\gamma$ absorption and  Na\,{\sc i} and CO bandhead emission  (Geballe et al. \cite{geballe}; their Fig. 6). Comparison to the spectrum of the YHG $\rho$ Cas presented in Yamamuro et al. (\cite{yamamuro}) suggest a similar classification for this star, although the near-IR properties (Table 1) imply a rather moderate luminosity in comparison to e.g. the YHG cohort of Westerlund 1 (Clark et al. \cite{clark05}).

\subsection{Interlopers and unclassifiable stars}

Given the crowded nature of the CMZ, the size of the IFUs and the selection criteria employed and it is inevitable 
that interlopers will be included in our survey. The most common contaminants were found to be cool, late-type stars in the line of sight which mirror the expected magnitudes and colours of early-type stars in the GC. Somewhat unexpectedly these were prevalent amongst targets selected
from the list of putative Pa$\alpha$ emitters of Dong et al. (\cite{dong11}).  Table 2 provides a list of such stars identified amongst our primary targets, along with those with either featureless spectra or insufficient S/N to attempt a classification. 
Where photometry was available, a combination of $(H-K)$ colour index and $K >10.5$mag implies that the cool stars are either foreground objects or descendants of rather low mass progenitors if found within the CMZ; hence we do not discuss these further. 
A large number of the IFUs contained one or more additional objects. In all relevant fields\footnote{Those surrounding stars [DWC2011] 15, 25, 34, 40, 43, 47, 48, 49, 57, 75,
78, 99, 105, 106, 135, 136, 143, and 147; S59, 64, 117, 123, 131, 136, and 183; MP13.}  the additional spectra were examined and found to be  either featureless or those of cool stars.

This leaves a handful of additional stars worthy of brief comment. Dong et al. (\cite{dong12}) classify both P38 and 140 as massive foreground 
objects on the basis of their near-IR colours. Based on our new spectrum, we reclassify the X-ray bright P38 slightly to O4 Ia. Mauerhan et al. (\cite{mauerhan10b}; their Fig. 3) give P140 as B0-2 Ia. However the spectrum presented by Geballe et al. 
(\cite{geballe}; their Fig. 6)  shows a pattern of emission (including Fe\,{\sc ii} and Na\,{\sc i}) and absorption (bluewards of Br$\gamma$) 
more reminiscent of a cool LBV candidate; intriguingly, comparison of these spectra implies variability between these observations (obtained in 2009 August and 2016 September respectively). The third object, P102, likewise has  near IR colours indicative of a foreground object, while the  double peaked Br$\gamma$ profile suggest that it is a classical Be star.

Lastly we turn to two objects for which we cannot yet advance a classification. The spectrum of P25 is featureless save for a single peaked 
Br$\gamma$ line of moderate intensity ($I/I_{cont}\sim1.4$). Unfortunately,  while the $K-$band measurement is consistent with the cohort of classical Be stars in the absence of colour information we may not infer a distance to the object and hence advance a classification.
 Finally P110 is of a comparable $K-$band magnitude and $(H-K)$ colour to the candidate classical Be stars (Sect. 3.7) and shows emission coincident with the wavelength of Br$\gamma$. However multiple apparently spurious  emission features of comparable strength are also present in the spectrum, casting some doubt as to the astrophysical origin for the putative Br$\gamma$ emission; hence we refrain from classifying the star at this time. 

\subsection{Synopsis}

In total we have observed 82 different primary targets, derived from the list of Pa$\alpha$ excess sources of Dong et al. 
(\cite{dong11})\footnote{For completeness we provide a list of Pa$\alpha$ excess sources remaining to be observed in Appendix A.}
 and previous reported candidates from the  literature. Of these the IFUs of 26 sources contained one or 
more additional objects - yielding spectra of $>100$ stars in total -  although none of these additional serendipitous sources were 
found to be massive stars. Of the primary targets a total of 31 were either cool foreground interlopers, displayed 
apparently featureless spectra  or were of insufficient S/N to attempt classification (Table 2).  Including one literature classification, three objects  were assessed as foreground massive stars while two further  were unclassifiable from our spectra (Sect. 3.9).

This left a total of 47 objects  which spectroscopy revealed to be massive stars. Of these 17 are new identifications, while the spectra of a further 19 stars allowed improved spectral classifications. The remaining 11 stars retained the same classification 
as previously reported.

Table 1 lists a total of 83 isolated massive stars - derived from our observations and the literature  - with near-IR photometry consistent with a location in the GC and for which classification  has proved possible\footnote{As noted in Sects. 3.3-3.7,
the diffuse, extended nature of the Quintuplet cluster makes it difficult to determine whether outlying massive stars are physically associated with it or not. Historically  the LBV G0.120-0.048 and WR stars WR102c (=qF353E; WN6), WR102ca (WC8-9) and CXOGC J174617.7-285007 (=MP14; WC8-9) have  been  included as probable cluster members (cf. Clark et al. \cite{clark18b}, \cite{clark20}),  despite their 
displacement from the central stellar concentration. As a consequence, and in order to avoid double counting, we do not include these in the census given here, but we highlight that their physical association with the cluster is as yet unproven.}. These comprise:

$\bullet$ Five mid-O and four late-O/early-B supergiants. One further object, XID \#947, was assigned a generic O classification with no indication of luminosity class by de Witt et al. (\cite{dewitt}). 

$\bullet$ Eight mid-to late-O hypergiants, 11  WNLha stars and one WN8-9ha/O6-7 Ia$^+$ hybrid. 

$\bullet$ Four WN9-11h stars, three early-B hypergiants and two hybrid early-B hypergiant/WNLh stars. Of these Geballe  et al. 
(\cite{geballe}) highlight the close proximity of 2MASS J17461292-2839001 (B1-3Ia$^+$) to the Quintuplet; we retain it here for 
completeness, although consider cluster membership highly  likely. 

$\bullet$ Eight objects with spectra characterised by strong narrow Br$\gamma$ and weak, low excitation metallic emission lines. We identify one  as a supergiant B[e] star on the basis of pronounced CO bandhead emission. We tentatively classify the remaining stars as either late-B hypergiants or cool phase LBVs; further photometric and 
spectroscopic monitoring being required to distinguish between these possibilities.

$\bullet$ One star, 2MASS J17444840-2902163, which we provisionally classify as a low-luminosity YHG and hence may extend the preceding cohort to lower temperatures.

$\bullet$ Eight WN5-7 stars, of which three are broad lined systems. Of these Mauerhan et al. (\cite{mauerhan10a}) note the proximity of P2
to the Quintuplet cluster. 

$\bullet$ 16 WC8-9 stars, including the featureless sources CXOGC J174645.2-281547, 2MASS J17432173-2951430, J17432988-2950074, and 2MASS J17460215-2857235 (Mauerhan et al. \cite{mauerhan10c}, Geballe et al. \cite{geballe}). 

$\bullet$ Six fainter stars with spectra dominated by strong Br$\gamma$ emission that we classify as classical Be stars.

$\bullet$ Three apparent (massive) YSOs and a further source displaying nebular emission associated with an UCH\,{\sc ii} region (noting that a considerably larger population of (massive) YSOs have been identified on the basis of IR-radio continuum and mid-IR spectroscopic observations; An et al. \cite{an}, Ginsburg et al. \cite{ginsburg18a}, Yusef-Zadeh et al. \cite{yz09}).

$\bullet$ One star, P57,  with a spectrum consistent with a classification as either a massive YSO or sgB[e] star.\newline

 Given the nature of the surveys employed for  target selection we emphasise that this census is highly likely to be incomplete. We discuss this limitation and the wider implications of this stellar population in the following sections.

\longtab{1}{
\begin{landscape}
\begin{longtable}{lccccclcc}
\caption{Census of massive stars in the sightline towards the GC}\\
\hline
\hline
ID         & RA      & Dec     & J     &    H     & K      & Aliases  & \multicolumn{2}{c}{Classification}  \\

[DWC2011]  & (h m s) & (d m s) & (mag) &    (mag) & (mag)  &          &  Old               & New            \\
\hline
\endfirsthead
\caption{continued.}\\
\hline
\hline
ID         & RA      & Dec     & J     &    H     & K      & Aliases  & \multicolumn{2}{c}{Classification}  \\

[DWC2011]  & (h m s) & (d m s) & (mag) &    (mag) & (mag)  &          &  Old               & New            \\
\hline
\endhead
\hline
\endfoot

\multicolumn{3}{l}{\underline{OB supergiants}}           &          &          &          &                                 &
&
\\
50  & 17 45 02.89 & -29 08 59.8 & 13.9$\pm$0.05  & 11.4$\pm$0.02  &  9.9$\pm$0.01  & CXOGC J174502.8-290859              & O9-B0 Ia &  -  
\\ 
95  & 17 45 59.44 & -28 52 50.7 & 14.9$\pm$0.02  & 12.1$\pm$0.01  & 10.5$\pm$0.02  &                                     &    -    & 
O9 Ia \\
     & 17 44 45.02 & -29 19 30.7 & 14.37          & 10.97          & 9.07           &  2MASS J17444501-2919307           & B2-3 Ia$^+$ 
& [B0-3 Ia] \\
    & 17 45 29.89 & -28 54 28.9 &   -            & 12.2           & 10.8           & S152             & - &  O4-5 Ia \\              
    & 17 45 30.32 & -28 52 07.0 &   -            & 13.4           & 11.9           & S73                                 &    -    & 
O4-5 Ia \\
    & 17 45 37.30 & -28 53 53.7 & 15.75$\pm$0.02 & 12.80$\pm$0.01 & 11.23$\pm$0.01 & CXOGC J174537.3-285354              & O9-B0 Ia& 
Similar \\
    & 17 46 28.31 & -28 39 20.5 & 16.99$\pm$0.07 & 13.36$\pm$0.03 & 11.49$\pm$0.04 & CXOGC J174628.2-283920  & O4-6 Ia      &  -  \\
    & 17 47 03.14 & -28 31 20.2 & 16.23$\pm$0.03 & 13.03$\pm$0.01 & 11.27$\pm$0.01 & CXOGC J174703.1-283119              & O4-6 Ia  &   -     
 \\
    & 17 47 25.41 & -28 25 23.0 &   -            & 13.37$\pm$0.01 & 11.30$\pm$0.04 & CXOGC J174725.3-282523              & O4-6 Ia  &   - \\

     &             &            &          &          &          &          &         &            \\
\multicolumn{3}{l}{\underline{WN7-9ha/O hypergiants}}  &          &          &          &         &    &       \\
15  & 17 46 03.21 & -28 48 58.4 &    -           & 13.2$\pm$0.05  & 11.7$\pm$0.03  &                         &   -         & O4-5 Ia$^+$ 
\\
22  & 17 45 53.40 & -28 49 36.9 & 15.4$\pm$0.02  & 12.5$\pm$0.02  & 11.0$\pm$0.01  & [MCD2010] 12            & WN8-9ha     &   -        
\\
23  & 17 46 10.01 & -28 55 32.4 & 15.0$\pm$0.02  & 12.3$\pm$0.02  & 10.8$\pm$0.02  & [MCD2010] 15            & WN8-9ha     & 
O6-7 Ia$^+$     
\\
35  & 17 45 28.62 & -28 56 05.0 & 14.5$\pm$0.02  & 11.5$\pm$0.02  &  9.7$\pm$0.03  & CXOGC J174528.6-285605, GCCR073  & O If$^+$    & WN8-9ha +neb 
\\
36  & 17 45 31.50 & -28 57 16.8 & 15.1$\pm$0.02  & 12.7$\pm$0.01  & 11.4$\pm$0.02  & [MCD2010] 6, CXOGC J174531.4-285716 & O4-6 Ia & 
O4-5 Ia$^+$ \\
75  & 17 46 17.10 & -28 51 31.5 & 15.0$\pm$0.03  & 12.1$\pm$0.12  & 10.5$\pm$0.02  & CXOGC J174617.0-285131  & O6 If$^+$   & WN8-9ha/O6-7 Ia$^+$ \\
77  & 17 46 17.54 & -28 53 03.5 & 15.0$\pm$0.02  & 12.1$\pm$0.01  & 10.5$\pm$0.02  & [MCD2010] 16            & WN8-9ha     &   -        
\\
96  & 17 45 48.57 & -28 50 05.7 & 17.6$\pm$0.10  & 13.3$\pm$0.02  & 11.0$\pm$0.02  & WR102a                  & WN8/Of      & WN7-8ha    
\\
97  & 17 45 47.72 & -28 50 49.2 & 15.3$\pm$0.19  & 12.3$\pm$0.09  & 10.7$\pm$0.06  &                         & O4-6 Ia$^+$ & O6-7 Ia$^+$ 
\\
100 & 17 45 42.32 & -28 52 47.1 & 14.7$\pm$0.02  & 11.7$\pm$0.02  & 10.1$\pm$0.03  & [MCD2010] 10            & O4-6 Ia$^+$ & 
O7-8 Ia$^+$ \\
107 & 17 45 39.34 & -28 53 21.1 & 14.7$\pm$0.02  & 11.8$\pm$0.01  & 10.2$\pm$0.01  &                         & O4-6 Ia$^+$ & 
O7-8 Ia$^+$ \\
111 & 17 45 36.12 & -28 56 38.7 & 15.6$\pm$0.03  & 12.3$\pm$0.01  & 10.4$\pm$0.02  & CXOGC J174536.1-285638, GCCR107  & WN8-9ha     &   -        
\\
114 & 17 45 32.78 & -28 56 16.6 & 14.7$\pm$0.02  & 12.1$\pm$0.01  & 10.7$\pm$0.01  & CXOGC J174532.7-285617, GCCR110  & O4-6 Ia & 
O4-5 Ia$^+$ \\
134 & 17 45 16.74 & -28 58 25.1 & 16.7$\pm$0.06  & 13.1$\pm$0.03  & 11.1$\pm$0.02  & CXOGC J174516.7-285824  & WN7-8ha     &   -        
\\

    & 17 46 18.79 & -28 49 48.4 &   -            & 12.5           & 11.1           & S131                    &  -  & O7-8 Ia$^+$ \\  
 
    & 17 46 20.87 & -28 46 58.8 &   -            & 14.1           & 12.5           & S120                    &  -  & WN8-9ha \\   

    & 17 46 56.36 & -28 32 32.3 &   -            & 13.74$\pm$0.05 & 11.24$\pm$0.02 & CXOGC J174656.3-283232  & WN8-9ha     &   -        
\\

    & 17 47 11.47 & -28 30 07.0 & 16.56$\pm0.06$ & 12.72$\pm$0.02 & 10.54$\pm$0.01 & CXOGC J174711.4-283006  & WN8-9ha     &   -        
\\

    & 17 47 12.25 & -28 31 21.6 & 17.06$\pm$0.07 & 13.07$\pm$0.01 & 10.78$\pm$0.02 & CXOGC J174712.2-283121  & WN7-8ha     &   -
\\

    & 17 47 13.03 & -28 27 08.2 &   -            & 14.22$\pm$0.01 & 11.86$\pm$0.01 & CXOGC J174713.0-282709  & WN7-8ha     &   -        
\\

   &             &           &          &          &          &                                 &         &            \\
\multicolumn{3}{l}{\underline{WN9-11h stars/early-B hypergiants}}           &          &          &          &
&         &            \\
19  & 17 45 48.61 & -28 49 42.2 &     -          & 13.6$\pm$0.03  & 11.1$\pm$0.01  & [MCD2010] 11          & WN8-9h    & WN9h(broad) \\

56  & 17 46 27.60 & -28 46 11.8 & 14.4$\pm$0.02  & 11.3$\pm$0.01  &  9.5$\pm$0.04  & [MCD2010] 18          & PCyg O Ia & WN11h      \\
98  & 17 45 41.27 & -28 51 47.7 & 14.8$\pm$0.02  & 11.6$\pm$0.01  &  9.9$\pm$0.01  & [MCD2010] 9           & O9-B0 If+ & B1-2 Ia$^+$/WNLh 
\\
103 & 17 46 01.65 & -28 55 15.3 & 13.4$\pm$0.01  & 10.7$\pm$0.01  &  9.1$\pm$0.05  & [MCD2010] 13          & PCyg OIa  & B0-1 Ia$^+$/WNLh 
\\
137 & 17 45 16.17 & -29 03 14.7 & 11.6$\pm$0.03  &  9.2$\pm$0.03  &  7.9$\pm$0.03  & CXOU J174516.1-290315, GCCR082 & Ofpe/WN9  & WN10h \\
    & 17 45 23.11 & -29 03 29.3 & 16.7$\pm$0.03  & 12.7$\pm$0.02  & 10.3$\pm$ 0.02 & SSTU J174523.11-290329.3 & B0-2 Ia & [B0-2 Ia$^+$] 
\\
    & 17 45 54.65 & -28 47 44.9 &   -            &  9.5           &  7.9           & S132                  & -         & B1-3 Ia$^+$ \\   
    & 17 46 12.93 & -28 49 00.2 & 13.79          & 10.53          &  8.85          & 2MASS J17461292-2849001  & -   &   [B1-3 Ia$^+$] \\  
    & 17 46 18.12 & -29 01 36.6 & 12.97$\pm$0.02 & 10.27$\pm$0.03 &  8.84$\pm$0.03 & WR102ka               &  WN10h    &  -    \\

\multicolumn{3}{l}{\underline{Cool BHGs/sgB[e] stars}}           &          &          &          &                                 &
&            \\
40  & 17 45 24.06 & -29 00 58.9 & 13.2$\pm$0.03  & 10.4$\pm$0.04 &  8.8$\pm$0.11 & [MCD2010] 5, 2MASS J17452405-2900589 & B0-2 Ia & 
sgB[e] \\
112 & 17 45 37.81 & -28 57 16.2 &    -           & 13.1$\pm$0.02 & 10.7$\pm$0.02 & 2MASS J17453782-2857161   & B[e] & late BHG/LBV?    \\
141 & 17 45 09.29 & -29 08 16.2 &    -           & 15.3$\pm$0.05 & 11.7$\pm$0.02 & 2MASS J17450929-2908164   &  -   & late BHG/LBV?    \\
    &  17 44 43.20 & -29 37 52.6 & 16.00          & 12.78         & 10.14         & 2MASS J17444319-2937526  &
uncl. & [late BHG/LBV?] \\
    &  17 44 55.38 & -29 41 28.5 & 16.72          & 12.63         & 10.14         & 2MASS J17445538-2941284  &
uncl. & [late BHG/LBV?] \\
    &  17 45 02.41 & -28 54 39.2 & 15.81          & 12.56         & 10.00         & 2MASS J17450241-2854392  &
uncl. & [late BHG/LBV?] \\
    &  17 47 09.40 & -28 49 23.6 & 12.66          & 10.8          &  9.46         & 2MASS J17470940-2849235  &
uncl. & [late BHG/LBV?] \\
    &  17 48 24.73 & -28 24 
31.3 & 15.86          & 12.12         &  9.54         & 2MASS J17482472-2824313  &
uncl. & [late BHG/LBV?] \\

    &             &           &          &          &          &                                 &         &            \\
\multicolumn{3}{l}{\underline{WN5-7 stars}}   &  &  &               &                                &        &         \\
2   & 17 46 23.81 & -28 48 10.8 & 16.3$\pm$0.05 & 13.3$\pm$0.01 & 11.4$\pm$0.02 & [MCD2010] 17                   & WN5b    & [WN5-6b]   \\
34  & 17 45 50.55 & -28 57 26.1 & 15.8$\pm$0.03 & 12.9$\pm$0.03 & 11.3$\pm$0.01 & [MCD2010] 19                   & WN5b    & WN6 \\
39  & 17 45 22.70 & -28 58 44.1 &   -           & 15.2$\pm$0.06 & 12.2$\pm$0.03 & CXOGC J174522.6-285844         & WN5-6b  & Similar \\
91  & 17 45 55.38 & -28 51 26.0 & 15.4$\pm$0.19 & 12.5$\pm$0.04 & 11.0$\pm$0.07 & CXOGC J174555.3-285126         & WN5-6b  & Similar \\
99  & 17 45 38.68 & -28 52 10.5 &  -            & 16.4$\pm$0.06 & 13.9$\pm$0.02 &                                &   -     & WN5-6  \\
109 & 17 45 50.61 & -28 59 19.6 & 15.4$\pm$0.02 & 12.5$\pm$0.01 & 10.9$\pm$0.01 & WR102b, CXOGC J174550.6-285919, GCCR085  &  WN7    & WN6-7 \\
147 & 17 45 08.98 & -29 12 17.9 & 15.1$\pm$0.03 & 12.6$\pm$0.02 & 11.1$\pm$0.04 & CXOGC J174508.9-291218         &  WN7    & WN6-7 \\
150 & 17 45 15.05 & -29 14 36.2 &   -           & 14.5$\pm$0.02 & 12.7$\pm$0.03 &                                &   -     &  WN5  \\ 

   &             &           &          &          &          &                                 &         &            \\
\multicolumn{3}{l}{\underline{WC8-9 stars}}           &          &          &          &                                 &    &   \\
28  & 17 45 57.76 & -28 54 45.8 & 16.3$\pm$0.03 & 13.4$\pm$0.01   & 11.4$\pm$0.02  & WR 101q                 & WC8-9       & Similar    \\
42  & 17 45 32.50 & -29 04 58.0 &    -          & 14.6$\pm$0.02   & 12.0$\pm$0.01  & [MCD2010] 8             & WC9         &   -        \\
49  & 17 45 21.89 & -29 11 59.6 &    -          & 14.7$\pm$0.03   & 12.7$\pm$0.03  & [MCD2010] 3             & WC9         & Similar    \\
53  & 17 45 07.07 & -29 12 00.6 & 16.8$\pm$0.07 & 13.5$\pm$0.02   & 11.1$\pm$0.03  & [MCD2010] 2             & WC9?d       & Similar    \\
94  & 17 46 02.59 & -28 54 14.0 & 16.4$\pm$0.04 & 13.5$\pm$0.02   & 11.6$\pm$0.02  & [MCD2010] 14            & WC9         &   -        \\
101 & 17 45 42.47 & -28 52 53.3 & 16.3$\pm$0.04 & 13.2$\pm$0.01   & 11.2$\pm$0.02  & WR101p                  & WC8-9       & Similar    \\
151 & 17 45 09.80 & -29 14 13.1 & 15.6$\pm$0.03 & 13.0$\pm$0.01   & 11.1$\pm$0.02  & [MCD2010] 4             & WC9?d       & Similar  \\  
    & 17 43 21.73 & -29 51 43.0 & 14.03         &  9.37           &  6.48          & 2MASS J17432173-2951430 & uncl.       & [WCLd]\\
    & 17 43 29.88 & -29 50 07.4 & 17.54         & 12.56           &  8.82          & 2MASS J17432988-2950074 & uncl.       & [WCLd]\\
    & 17 44 37.35 & -29 27 55.7 & 17.56         & 13.91           & 10.29          & 2MASS J17443734-2927557 & WCLd        &   - \\
    & 17 44 40.83 & -29 26 55.1 & 16.76         & 12.81           &  9.40          & 2MASS J17444083-2926550 & WCLd        &   - \\
    & 17 45 04.84 & -29 11 46.5 & 15.02         & 11.63           &  9.04          & 2MASS J17450483-2911464 & WCLd        &   - \\
    & 17 45 19.17 & -29 03 22.0 & 17.10         & 13.30$\pm$0.05  & 10.40$\pm$0.05 & CXOGC J174519.1-290321, MP13 &  WC9   & Similar    \\
    & 17 46 02.16 & -28 57 23.5 & 14.87         & 11.43           &  8.08          & 2MASS J17460215-2857235 & uncl.       & [WCLd] \\
    & 17 46 32.20 & -28 44 54.6 & 17.14         & 12.86           &  9.21          & 2MASS J17463219-2844546 & WCLd        &   - \\
    & 17 46 45.25 & -28 15 47.7 & 15.39         &  9.96$\pm$0.03  &  7.18          & CXOGC J174645.2-281547  & WCLd        &   -        \\

   &             &           &          &          &          &                                 &         &            \\
\multicolumn{3}{l}{\underline{Classical Be stars}}           &          &          &          &                        &     &   \\
105 & 17 45 52.97 & -28 55 36.9 & 17.69$\pm$0.16 & 14.75$\pm$0.05 & 13.08$\pm$0.05 & CXOGC J174552.9-285537, XID \#3275 & Be star & Similar \\
106 & 17 45 52.26 & -28 56 46.1 &       -        & 16.1$\pm$0.08  &  14.5$\pm$0.13 &                                 &    -    & Be star \\ 
108 & 17 45 47.04 & -28 56 46.0 &       -        & 15.5$\pm$0.07  &  14.0$\pm$0.09 &                                 &    -    & Be star \\
135 & 17 45 17.76 & -28 58 19.9 &       -        & 16.1$\pm$0.10  & 14.1$\pm$00.06 &                                 &    -    & Be star \\
\multicolumn{3}{l}{\underline{Classical Be stars (cont.)}}     &          &          &          &                        &     &   \\  
    & 17 45 52.16 & -28 55 01.5 &       -        & 17.4           & 14.6           &                          S69    &    -    & Be star \\
    & 17 46 12.95 & -28 47 16.3 &       -        & 14.6           & 13.1           &        S124                     &    -    & Be star \\

  &             &           &          &          &          &                                 &         &            \\
\multicolumn{4}{l}{\underline{Miscellaneous and  uncertain classification}}              &          &          &                                 &         &      \\
57  & 17 46 29.90 & -28 46 39.9 & -              & 14.7$\pm$0.02 & 12.6$\pm$0.02 &                          &   -   & MYSO/sgB[e] \\
58  & 17 46 31.85 & -28 46 47.1 &  14.3$\pm$0.02 & 12.6$\pm$0.01 & 11.6$\pm$0.01 &                          &  -    & MYSO \\       
    & 17 43 10.02 & -29 51 46.1 & 17.01          & 13.91         & 11.58         & 2MASS J17431001-2951460 & uncl.  & [MYSO] \\
    & 17 44 48.40 & -29 02 16.4 & 13.81          & 12.08         & 10.34         & 2MASS J17444840-2902163 & uncl.  & [YHG]  \\ 
    & 17 47 09.22 & -28 46 16.2 & 14.21          & 12.58         & 10.18         & 2MASS J17470921-2846161 & uncl.  & [MYSO] \\ 
    & 17 45 28.88 & -28 57 26.4 & 16.23$\pm$0.06 & 13.29$\pm$0.03  & 11.61$\pm$0.03 & XID \#947, GCCR072 & O star?  &   -   \\
    & 17 45 51.54 & -29 00 23.2 &    -           & 14.88         & 11.43         & [HJB85] Sgr A-D         & LBV/ucH\,{\sc ii} 
& ucH\,{\sc ii} \\
  &             &           &          &          &          &                                 &         &            \\
\multicolumn{4}{l}{\underline{Foreground (($H-K)\leq1$) and unclassifiable}}                      &          &          &                                 &         &            
\\
25  & 17 45 58.31 & -28 52 20.0 &     -         &      -       & 13.4$\pm$0.06  &                                   &   -     & Uncl. \\ 
38  & 17 45 37.99 & -29 01 34.5 & 11.1$\pm$0.05 & 9.7$\pm$0.05 & 8.9$\pm$0.13 & [MCD2010] 7, CXOGC J174537.9-290134 & O4-6 Ia & O4 Ia \\
102 & 17 45 44.04 & -28 53 16.8 & 15.0$\pm$0.02 &13.3$\pm$0.03 &12.4$\pm$0.04 &                                     &   -     & Be star \\
110 & 17 45 48.94 & -28 59 01.8 &     -         &15.7$\pm$0.04 &14.3$\pm$0.03 &                                     &   -     & Uncl. \\     
140 & 17 44 59.46 & -29 05 25.9 &  8.8$\pm$0.03 & 7.7$\pm$0.04 & 7.0$\pm$0.03 & [MCD2010] 1, 2MASS J17445945-2905258 &  B0-2 Ia & [cool 
LBV] \\

\end{longtable}
{Summary of known and candidate isolated massive stars in the sightline to the GC.  Columns 1 and 7 identify the source and provide relevant aliases, columns 2 and 3 provide 
co-ordinates and 4-6 provide broadband near-IR photometry. Finally column 
8 provides, where available,  historical classifications for these stars 
and column 9 our new or revised classifications from this work. We highlight that a 
subset of such new classifications are given in square brackets - these 
derive from our reappraisal of published data. Where available we adopt 
the naming convention of Dong et al. 
(\cite{dong11}; [DWC2011] {\em xxx}) for their primary list of 
Pa$\alpha$ excess sources (column 1), noting  that S69, 73, 120, 124, 131, 132, and 152 derive from 
their secondary list. Additional identifications derive from  Mauerhan et al. (\cite{mauerhan10a}; 
CXOGC sources), Mauerhan et al.  (\cite{mauerhan10c}; [MCD2010]{\em xx}) 
de Witt et al. (\cite{dewitt} XID {\em xxx}), Geballe et al. (\cite{geballe}, 2MASS sources) and Zhao et al. (\cite{zhao20}; GCCR{\em xxx}). In all cases 
photometry (and associated errors where available) and classifications were obtained from these 
works, supplemented by Homeier et al. (\cite{homeier}), Mauerhan et al. 
(\cite{mauerhan07}), Clark et al. (\cite{clark09b}), Oskinova et al.
(\cite{oskinova}), de Witt et al. (\cite{dewitt}) and Dong et al. (\cite{dong15}) where required.}
\end{landscape}
}

\begin{table}
\caption{Non massive star interlopers projected onto the CMZ}
\begin{center}
\begin{tabular}{llll}
\hline
\hline
[DWC2011]  &  RA (J2000) & Dec (J2000) & Appearance \\
\hline
3     & 17 46 03.60 & -28 47 09.8 & Cool \\
16    & 17 45 54.85 & -28 47 13.2 & Cool \\ 
29    & 17 46 09.60 & -28 57 13.5 & Cool \\
30    & 17 46 08.10 & -28 58 24.2 & Cool \\
31    & 17 46 06.27 & -28 59 14.9 & Cool \\
43    & 17 45 13.95 & -29 04 38.2 & Cool \\
47    & 17 45 15.05 & -29 09 08.3 & Featureless \\
48    & 17 45 22.68 & -29 10 56.5 & Featureless \\
78    & 17 45 48.22 & -28 47 26.0 & Cool \\
104   & 17 46 07.30 & -28 57 17.5 & Cool \\
113   & 17 45 33.69 & -28 57 49.7 & Cool \\
133   & 17 45 36.86 & -29 01 17.5 & Featureless \\
136   & 17 45 28.22 & -29 03 27.1 & Cool \\
142   & 17 45 19.45 & -29 10 33.9 & Low S/N \\
143   & 17 45 23.91 & -29 10 24.7 & Low S/N \\
144   & 17 45 14.22 & -29 11 41.7 & Low S/N \\
(S47) & 17 46 31.07 & -28 46 14.2 & Cool \\
(S52) & 17 46 24.30 & -28 46 53.1 & Cool \\
(S59) & 17 46 01.64 & -28 49 00.0 & Cool \\ 
(S64) & 17 46 05.02 & -28 52 06.7 & Cool \\
(S78) & 17 45 53.36 & -29 00 49.1 & Low S/N \\
(S113)& 17 44 55.32 & -29 11 54.8 & Cool \\
(S117)& 17 46 26.56 & -28 46 26.4 & Cool \\ 
(S119)& 17 46 32.72 & -28 46 08.1 & Cool \\
(S123)& 17 46 29.95 & -28 49 32.2 & Low S/N \\
(S136)& 17 46 01.64 & -28 51 10.3 & Featureless \\
(S139)& 17 46 02.44 & -28 53 13.7 & Cool \\
(S141)& 17 46 06.14 & -28 54 41.4 & Low S/N \\
(S146)& 17 46 07.43 & -28 59 18.5 & Low S/N \\
(S153)& 17 45 35.44 & -28 55 12.3 & Low S/N \\
(S183)& 17 45 04.24 & -29 09 26.7 & Cool \\
\hline
\end{tabular}
\end{center}
{Stars given in parentheses with an `S{\em xxx}' designation are from the secondary list of Pa$\alpha$ 
emitters in Dong et al. (\cite{dong11}).}
\end{table}

\section{Discussion} 

The results described in the preceding section reveal that a large number of isolated and potentially very massive stars of diverse nature appear distributed through the CMZ. These include very rare phases -  such as LBVs and sgB[e] stars - that are thought to play an important   role in the lifecycle  of single and binary stars. As such characterising  this population,  including observational biases,  will be critical to further  understanding the evolution of such extreme objects as well as the wider ecology and star formation history  of the circumnuclear region of the Galaxy.

\subsection{Survey completeness and  biases}

Centred on Sgr A$^*$, the HST Pa$\alpha$ excess survey has the smallest (asymmetric) footprint of those utilised  in our study
($39\times15$ arcmin or $93.6\times36$ pc at a distance of 8kpc; Wang et al. \cite{wang}, Dong et al. \cite{dong11}) and is expected to detect early-type stars via emission from their ionised wind. As such detection probability will be a function of both stellar mass and evolutionary phase, with mass loss rates greater for more massive stars and wind densities increasing through the stellar 
lifecycle\footnote{Subject to the caveat that stars in a cool yellow hypergiant/red supergiant phase will be undetectable unless their winds are subject to external ionsiation (cf. Westerlund 1; Dougherty et al. \cite{dougherty10}, Fenech et al. \cite{fenech})}. Since all three young massive clusters are within the survey footprint we may utilise their well defined stellar populations to empirically determine its sensitivity to spectral type.

 All 13 WNLha stars and 5/7 of the  O hypergiants within the Arches are detected via their Pa$\alpha$ excess, but only 3/30 of the mid-O supergiants and none of the $>50$ O5-9 stars of luminosity class III to V, although source blending/confusion may compromise identification given the compact nature of the cluster (Dong et al. \cite{dong11}, Clark et al. \cite{clark18a}, \cite{clark19a}). 
In the Quintuplet - which is less compact and so presumably less prone to blending - all three  LBVs and 8/10 of the early-B HGs/WNLh are detected (with the two missing examples having the weakest He\,{\sc i} 2.059 emission and no trace of
BrG emission), along with  the sole WN6, 7/16 of the WC stars and all 5 of the candidate 
`blue stragglers' of spectral types WN8-9ha and O7-8Ia$^+$. However, as with the  Arches only 1/23 of the O7-B0 Ia supergiants are 
detected. Although apparently of lower luminosity (Martins et al. \cite{martins07}), the same pattern is repeated for stars within the Galactic Centre cluster, with 25/33 of the Wolf-Rayet cohort detected but only one of the 26 OB supergiants present.

Consistent with our findings for the isolated stellar cohort (Tables 1 and 3), the cluster detection demographics imply that even stars as extreme as O4 supergiants may be routinely undetectable via their excess Pa$\alpha$ emission. 
This is all the more striking since such objects likely derive from very massive stars (M$_{\rm init}{\gtrsim}40M_{\odot}$; Groh et al. 
\cite{groh14}, Martins \& Palacios \cite{martins17}, Clark et al. \cite{clark18a}). 

The Chandra GC survey of Muno et al. (\cite{muno09}) covers a larger field than the preceding study 
($2^{\rm o}\times0.8^{\rm o}$  or $\sim280\times112$ pc at a distance of 8kpc). In total Mauerhan (\cite{mauerhan09}) suggest $\sim100-300$ X-ray sources are coincident with near-IR sources with $K<15.6$; consistent with both late-type giants and massive stars with luminosities equal to, or greater than,  early-B dwarfs. Due to the high column density towards the GC  we would not anticipate detecting the comparatively soft X-ray emission expected from shocks embedded in the winds of single stars, suggesting the latter cohort comprise colliding wind binaries (CWB) and, if present, accreting high-mass X-ray binaries. 

Unfortunately, the X-ray emission from CWBs appears a sensitive function of a number of physical  properties 
(component masses, wind velocities and mass loss rates, orbital separation and eccentricity) to the extent that such systems can show no enhancement in X-ray emission over that expected from a single star, enhanced but soft X-ray emission or excess hard X-ray emission 
(cf. Westerlund 1; Clark et al. \cite{clark19b}). Given the first two scenarios would not in general lead to detectable sources we assume this survey is likely incomplete even for massive CWBs. Empirically, Mauerhan et al. (\cite{mauerhan10c}) identified 18 massive stars from a sample of 52 near-IR bright ($K<12$) matches to hard X-ray sources. However a number of targets were selected  on the basis of their proximity to mid-IR structures indicative of the presence of massive stars (i.e. wind-blown bubbles and bow shocks),  precluding a statistical analysis of the success rate deriving from the application of such observational criteria. 
Nevertheless, the distribution of the spectral types of the X-ray bright massive stars essentially mirrors that derived from the Pa$\alpha$ survey (Table 1 and Mauerhan et al. \cite{mauerhan10c}); such a detection bias is unsurprising since both  surveys are expected to be  sensitive to stars supporting dense, high velocity winds.

The photometrically selected survey of Geballe et al. (\cite{geballe}) has a footprint  intermediate between the preceding studies  
($2.4^{\rm o}\times0.6^{\rm o}$  or $\sim336\times84$pc at a distance of 8kpc). Designed to select candidate for follow-up studies of 
warm diffuse gas it employed a mid-IR cut ([3.6]$<8$) that is not optimised for identifying massive stars.
This is evident in the detection rate, with $\sim32$ candidate massive stars from over 500 spectroscopically surveyed; these being 
strongly biased towards sources potentially associated with circumstellar dust such as cool LBVs/sgB[e]  and WCLd stars\footnote{Intriguingly the sample of WCLd stars returned shows no overlap with that selected via Pa$\alpha$ excess; presumably emission from the warm dust that allows their photometric detection  dilutes emission features in the spectrum to the point at which they cannot meet the  detection threshold for the survey of Dong et al. (\cite{dong11}).}. No WN5-7 or Be stars  were detected, and the three O hypergiants and WNLha stars identified were coincident with extended mid-IR nebulae, suggesting they were still associated with their natal material or bow shocks (P35 and 114 respectively; Dong et al. \cite{dong17}).

In conclusion the combination of different methodologies  (and in some cases subjective criteria) 
means that we are unable to provide robust quantitative estimates for the detection thresholds and hence level of incompleteness of the surveys informing our target selection. However it seem likely that all three 
are insensitive to a large number of stars over a wide range of initial masses and evolutionary phases. Specifically, one would only expect to detect comparatively low mass objects ($M_{init}\sim8-25M_{\odot}$; Groh et al. \cite{groh13}) at the  end their lives, via dust emission in an RSG phase or via a Pa$\alpha$ excess in a Be star episode. Stars above this threshold
 are expected to loop back to higher temperatures and  hence one might also anticipate detecting them in a  LBV/BHG or WR phase. The same is expected at still higher masses ($M_{init}\gtrsim40M_{\odot}$) -  a regime in which stars  remain at high  temperatures throughout their lives. 
However, evolutionary simulations suggest that such stars will only support winds of sufficient density to permit detection after a considerable  proportion of their H-burning lifetime has elapsed. For example the simulations of Groh et al. (\cite{groh14}) show that a M$_{\rm init}{\sim}60M_{\odot}$ star has a lifetime of $\sim4.0$Myr, but only reaches the early-B hypergiant phase - which observations of the Quintuplet suggest are detectable via their Pa$\alpha$ excess - after $\sim3.3$Myr, potentially leaving it undetectable by such a survey for ${\gtrsim}80$\% of its life.

Informed by the stellar population of  Westerlund 1, one might expect binarity to aid detection via all survey methodologies. Mass loss during the active interaction potentially leads to the formation of sgB[e] stars (cf. Wd1-9; Clark et al. \cite{clark13}; Kastner et al. \cite{kastner}) which should be readily 
detectable as Pa$\alpha$ excess sources (cf. P40 and 57). Likewise stripped primaries entering a  (proto-)WR phase 
(cf. Wd1-5; Clark et al. \cite{clark14}) and mass-gainers/merger products (cf. Wd1-27 and 30a; Clark et al. \cite{clark19c}) both support pronounced emission line spectra which are absent in their late-O supergiant progenitors. Moreover, dust production in CWBs containing WC stars enhances the likelihood of their detection at mid-IR wavelengths, while X-ray emission via wind collision or, more rarely, accretion onto a compact companion may render identifiable otherwise undetectable stars.

\subsection{The origin and spatial distribution of  isolated massive  stars}
 
 Various authors have suggested that massive stars may form in comparative `isolation'  in regions of low molecular and, subsequently, stellar density (de Wit \cite{dewit04}, \cite{dewit05}, Parker \& Goodwin \cite{parker}). Indeed, observations  of the Cyg OB2 association (Wright et al. \cite{wright14}, \cite{wright16}) the 30 Dor star forming region (Bressert et al. \cite{bressert}, Schneider et al. \cite{schneider18}) and the Small Magellanic Cloud
(Lamb et al. \cite{lamb}) are consistent with such an hypothesis.
However, determining the origin of isolated stars is notoriously difficult;  the WN5h  star VFTS 682 - which closely resembles the  WNLha stars distributed through the CMZ - being a case in point. Bestenlehner et al. (\cite{best}) highlight its location in the outskirts of 30 Dor and  the  lack of an associated stellar aggregate as consistent with its formation in isolation, but are unable to exclude  an origin in -  and subsequent ejection from -  the young massive cluster R136, some 29pc distant  (which is itself an analogue of the Arches; Crowther et al. \cite{crowther16}). Therefore before addressing the distribution of isolated stars through the CMZ it is  instructive to consider the physical mechanisms that may redistribute stars from cluster to field.

\subsubsection{Stellar redistribution across the CMZ}

Two physical mechanisms are thought to give rise to the majority of runaway stars in the Galactic disc; ejection via SN explosions in binaries  (Blaauw \cite{blaauw}) or dynamical interaction (Poveda et al. \cite{poveda}, Banerjee et al. \cite{banerjee}, Fujii \& Portegies Zwart \cite{fujii}). Considerable effort has been invested in determining the magnitudes of  SN kicks and their effects on the survivability and motion of binary systems (e.g Renzo et al. \cite{renzo} and refs. therein). With $v\sim90$kms$^{-1}$, the high mass X-ray binary  Vela X-1 (B0 Ib + neutron star) suggests that a considerable velocity may be imparted in at least some cases (Kaper et al. \cite{kaper97}). While this channel appears inapplicable for the Arches given its age 
(2-3Myr; Clark et al. \cite{clark18a}), it is likely viable for both the Quintuplet and Galactic Centre clusters ($\sim3-3.6$Myr and $\sim4-8$Myr respectively; Clark et al. \cite{clark18b}, Paumard et al. \cite{paumard}). Indeed the presence of the magnetar SGR J1745-29
within $\sim3$" of Sgr A* (Kennea et al. \cite{kennea}, Mori et al. \cite{mori}) and the young pulsar J1746-2850I within 2' of the Quintuplet (Deneva et al. \cite{deneva}) are suggestive of ongoing SNe activity in these regions.

Turning to dynamical ejection and one example of a  very massive and high velocity runaway is VFTS 16 ($M\sim100M_{\odot}$, $v\sim112$kms$^{-1}$),   which Lennon et al. (\cite{lennon}) demonstrate to have a  proper motion consistent with an origin in the LMC  cluster R136; an aggregate that is too young to host SNe at this time (Crowther et al. \cite{crowther16}).
 This is of particular interest since R136 appears similar to the Arches in terms of its youth, stellar density and masses of constituent stars (Clark et al. \cite{clark18a}, \cite{clark19a}); suggesting that similar high velocity runaways might be expected to  originate from the latter cluster. 

On larger physical scales, the expulsion of residual gas from compact clusters via stellar feedback has long been posited as a mechanism for driving their expansion and, in some cases,  destruction - with the now supervirial velocities of the constituent stars dispersing them into the wider field (e.g. Goodwin \& Bastian \cite{gb}; see also Park et al. \cite{park18} for a discussion of this effect in the context of the GC). Moreover, the tidal stripping and the eventual disruption of clusters may also distribute massive stars through  the GC. Simulations of this phenomenon for both the Arches and Quintuplet suggest that tidal arms of several tens of parsecs may result from this process after only a few Myr, although the extent of such structures is a sensitive function of both cluster age and distance from Sgr A$^*$ (Habibi et al. \cite{habibi}, Park et al. \cite{park}). By comparison, assuming that both dynamical interactions and SNe kicks may generate runaway velocities of up to $\sim10^2$kms$^{-1}$ such stars may be displaced from their natal clusters by up to $\sim100$pc ($\sim0.7^{\rm o}$ at 8kpc) within $10^6$yr - comfortably less than the age of either the Arches or Quintuplet. Combined with the bulk orbital motion of both aggregates ($232\pm30$kms$^{-1}$ and $167\pm15$kms$^{-1}$ respectively; Stolte et al. \cite{stolte08}, \cite{stolte14}) one may anticipate a combination of these three processes potentially  stripping massive  stars from their natal clusters and distributing them  across a significant fraction of the CMZ.

\subsubsection{The  distribution of massive stars across the CMZ}

\begin{figure*}
\includegraphics[width=20cm,angle=-0]{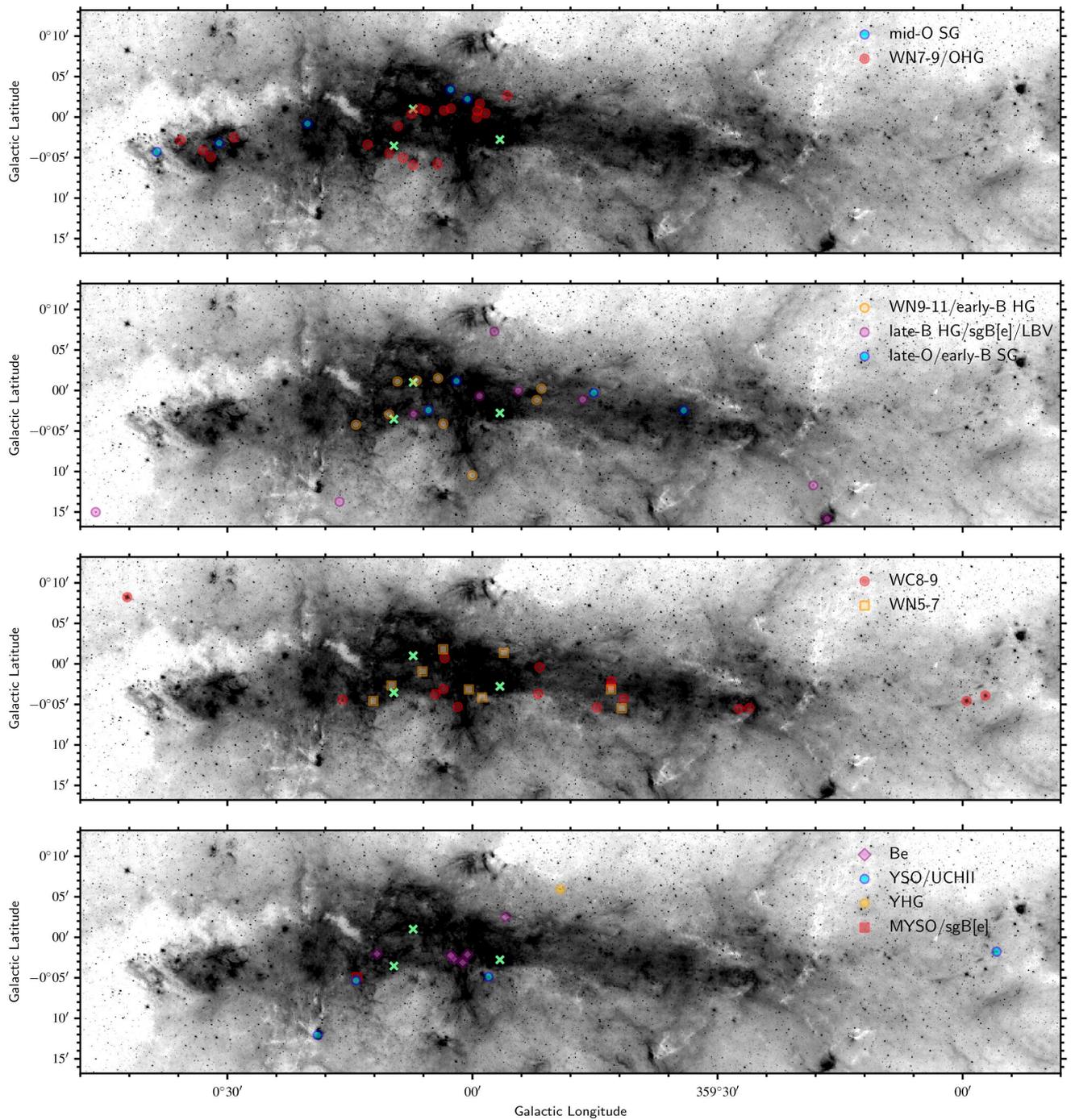}
\caption{Location of different classes of massive stars superimposed on a greyscale representation of {\em Spitzer} 8$\mu$m
continuum data. The location of the Galactic Centre, Arches and Quintuplet clusters are given by green crosses. Given their displacement 
from the nominal location of the Quintuplet cluster we have also plotted the positions of the LBV G0.120-0.048 and the WN6 star qF353E
even though  they are not included in the census on Table 1. Each panel covers $\sim208\times70$pc at a distance of 8kpc.
}
\end{figure*}

We show the locations of the isolated massive stars across the GC in Fig. 14.  In constructing  this plot we omitted 
both OB dwarfs which, to date, have solely been identified within the Arches and Galactic Centre clusters (Clark et al. \cite{clark19a}, Paumard et al. \cite{paumard}) 
and red supergiants, which are present (e.g. Wollman et al. \cite{wollman}, Cunha et al. \cite{cunha}, Liermann et al. \cite{liermann12}) but difficult to distinguish from cool interlopers with the data to hand.

It is immediately apparent that there is a  large overdensity  of stars in the inner regions of the CMZ; we consider it likely that this is in part due to the limited footprint of the  Pa$\alpha$ survey (running from $\sim +0^{\circ} 14'$ to $\sim 359^{\circ} 38'$; Wang et al. \cite{wang}) which drove placement of the   KMOS IFUs.
Given such observational biases and the potential redistribution of stars from their birth sites, we are limited in the  conclusions that may be  drawn from these data, especially for individual objects.

Nevertheless, thanks to the sample size we are able to plot massive, post-main sequence stars as a function of spectral type which,  informed by cluster demographics,  we may use as a proxy for stellar age. Comparison to the Arches cluster suggests that the mid-O super-/hypergiants and WNLha stars plotted in the uppermost panel of Fig. 14 are likely to be the youngest and most massive cohort present ($<3$Myr and $M_{\rm init}\gtrsim40M_{\odot}$; cf. Clark et al. \cite{clark18a}). A cursory examination reveals this cohort is observed at low Galactic latitudes, as might be expected for young objects. More intriguingly,
the entire population appears to be distributed at positive galactic longitudes, with no examples significantly westward of Sgr A$^*$ (the right half of the panel in Fig. 14). Given that the entire  region is covered by both the Pa$\alpha$ and X-ray surveys, which are sensitive to such stars, it appears difficult to attribute this to observational bias. 

Historically it had been thought that the bulk of the dense molecular gas  within the CMZ was located at positive galactic longitudes (so eastwards of Sgr A$^*$ in Fig. 14) while the majority  of young stars identified by mid-IR observations were instead found  at negative longitudes (Morris \& Serabyn \cite{MS}, Longmore \& Kruijssen \cite{longmore}). However, the  distribution of mid-O super-/hypergiants and WNLha stars shows the situation is less clear cut,  with a rich population of young, very massive stars coincident with active star forming regions. Habibi et al. (\cite{habibi}) suggested that a noteworthy association of six  stars\footnote{CXOGC J174656.3-283232 (WN8-9ha), CXOGC J174703.1-283119 (O4-6 Ia), CXOGC J174711.4-283006 (WN8-9ha), CXOGC J174712.2-283121 (WN7-8ha), CXOGC
J174713.0-282709 (WN7-8ha) and CXOGC J174725.3-282523 (O4-6 Ia).} co-located with the Sgr B complex (RA  $\sim0^{\circ} 27' \rightarrow 0^{\circ} 42'$, $\delta \sim 0^{\circ} 0' \rightarrow -0^{\circ} 8'$)   could originate in a tidal arm associated with the Quintuplet. However the  cluster age and membership profile precludes  such an origin for these objects - and indeed for any of the mid-O  super-/hypergiants and WNLha stars within the GC distributed via any dispersal mechanism (cf. Clark et al. \cite{clark18b}). The only cluster young enough to serve as a viable birthsite is the Arches (Clark et al. \cite{clark18a}); however predictions for the geometry of a putative tidal tail suggest that it would not intersect with Sgr B (Habibi et al. \cite{habibi}). Therefore, given their apparent co-evality and proximity to one another we consider it likely that the sextet form a distinct, physically related  group; we discuss this putative aggregate further in Sect. 4.2.3.

Conversely, as discussed in Clark et al. (\cite{clark19a}) the mid-O supergiants and WNLha stars adjacent to the  Arches cluster could have originated there and been ejected via dynamical interactions or tidal stripping. A further grouping of four WNLha stars and mid-O hypergiants with small angular separations from one another are located north east of Sgr A$^*$ (RA  $\sim359^{\circ} 59'$,  $\delta \sim 0^{\circ} 0'$); we revisit this possible assemblage in Sect. 4.2.3.

We plot the locations of the  late-O to late-B  super-/hypergiants, LBVs and WN9-11h stars - so called `transitional' objects  - in panel two and the H-depleted WN5-7 and WC8-9 stars in panel three of Fig. 14. Comparison to the populations of the Quintuplet and GC suggest they derive from an older stellar  population ($\sim3-8$Myr) than the mid-O hypergiants and WNLha stars and one that is likely to be more heterogeneous  in terms of initial mass ($M_{init} \gtrsim30M_{\odot}$; cf.  Paumard et al. \cite{paumard}, Martins et al. \cite{martins07}, Clark et al. \cite{clark18b}). While both cohorts are again centrally concentrated there appears less evidence for segregation as a function of galactic longitude in comparison to the younger stars. We note, however, that there is an absence of older stars coincident with Sgr B; on this basis one might speculate that  active star formation in this region  has proceded for $\lesssim3$Myr. 

Mindful of observation biases and a comparatively small sample size, a greater number of the transitional stars and H-depleted WRs appear to be seen in relative isolation at more extreme galactic latitudes and longitudes than found for the younger cohort of mid-O hypergiants and WNLha stars.   
If these trends are borne out by future observations one might anticipate that their greater ages have allowed  runaways to move further from their birthsites (noting that some stars, such as the WN10h star WR102ka, appear to have formed in isolation; Oskinova et al. \cite{oskinova}). Indeed, there are a large number of stars seen in close proximity to both the Quintuplet and Galactic Centre clusters that could plausibly have originated in those aggregates; follow up proper motion observations of these would be of considerable interest (cf. Sect. 4.2.3).

Finally, in this regard we note that a substantial number of isolated stars appear to demonstrate multiwavelength properties suggestive of a binary nature. Specifically, given the high column density towards
the Galactic centre, the 24 X-ray detections must by necessity be hard, luminous sources indicative of emission from a wind collision zone (Mauerhan et al. \cite{mauerhan10a}, Table 1). Similarly, such structures are also thought to mediate the production of hot dust, yielding a further six candidate binaries amongst the WCL stars (Table 1)\footnote{The remaining dusty source, CXOGC J174645.2-281547
has already been  counted as an X-ray source.}. While such systems could arise from in situ formation, the tidal disruption of a cluster or dynamical ejection from such a site 
(Oh \& Kroupa \cite{oh}) formation via a  SN kick would appear disfavoured; future studies to compare the binary properties of isolated stars and those within clusters would be of particular 
interest in order to elucidate their formation pathways.

\subsubsection{Clustered star formation within the CMZ}

Simple visual inspection of the CMZ fails to reveal any comparably massive clusters to the Arches or Quintuplet. This absence is unlikely to be driven by stellar evolution; although the most massive members will be lost to SNe the remnant population of an ageing cluster will increasingly become dominated by IR-bright RSGs. 
For example, if placed in the GC, and assuming a representative  extinction of $A_K{\sim}3$, the ten cool super-/hypergiants within Westerlund 1 would span $K\sim5-8$  and lie within a nominal cluster diameter of $\sim1.25$arcmin (cf. Borgman et al. \cite{borgman}); the respective parameters for  RSGC1-4 being broadly equivalent (Figer et al. \cite{figer06}, Davies et al. 
\cite{davies07}, \cite{davies08}, Clark et al. \cite{clark09a}, Negueruela et al. \cite{negueruela10}).
 Such apparent magnitudes are comparable to those of the  WCLd stars that dominate the appearance of the Quintuplet - which is also of similar  angular extent -  suggesting that if older stellar aggregates were present within the GC they should be readily identifiable in existing surveys. Since the timescale for cluster disruption due to tidal forces is expected to be  short  (e.g. 5-10Myr; Portegies-Zwart et al. \cite{pz}) we might anticipate that their  absence is due to  dissolution; although individually luminous, if their constituent stars were distributed through the CMZ they would be difficult to distinguish from foreground objects.
 
We may ask whether there is any evidence for such a process occurring within the GC. Located within 6" of Sgr A$^*$, GC IRS 13E is a compact (d$\sim0.5$ arcsec) group of three co-moving massive stars (Fritz et al. \cite{fritz}) that has been suggested to be the tidally stripped remnant of a young massive cluster and hence provides a possible template for the appearance of  such physical systems.  Although the angular resolution of our KMOS observations is lower than those used to characterise IRS 13E, the field of view of each IFU ($2.8\times2.8$ arcsec) is ideally suited to the detection of  such compact stellar aggregates. Despite the emplacement of 82 IFUs across the GC and the presence of multiple stars within $\sim32$\% of these, no further candidates were found in our observations (Sect. 3.9). 

At larger angular scales the spatial coincidence  of a number of very massive, apparently co-eval  stars with the Sgr B star forming complex is of considerable interest (Fig. 14 and Sect. 4.2.2). Sgr B comprises three distinct 
regions that appear to be  embedded within the same molecular cloud; Sgr B1, B2 and G0.6-0.0, which bridges the gap between the two former zones. Observations from mid-IR to radio wavelengths suggest that Sgr B2 is actively forming stars at this time (e.g. Ginsburg et al. \cite{ginsburg18a}),  while  Sgr B1 appears older, with bright rimmed and shell-like structures suggestive of  the action of 
stellar winds (Mehringer et al. \cite{mehringer}, Hankins et al. \cite{hankins}).    Simpson et al. (\cite{simpson}) infer $T_{\rm eff}\sim32-35$kK  for the ionising sources of Sgr B1; directly comparable to the temperatures expected for the WNLha and mid-O super-/hypergiants co-located with it (Footnote 14; Martins et al. \cite{martins08}, Lohr et al. \cite{lohr}). Moreover, such stars  generate the powerful stellar winds and ionising radiation fields required to sculpt and disperse  the molecular material associated with Sgr B1; further strengthening the possibility of a physical association between the two.

 Simpson et al. (\cite{simpson}) suggest that the exciting sources of Sgr B1 did not originate there, but instead are only now impinging on the  molecular cloud by virtue of their orbital motion.  While proper motion studies will be required to assess whether they were born in situ or not, either eventuality implies that the  WNHLha and mid-O super-/hypergiants form a genuine, co-eval  physical association. This is particularly interesting since, lying  outside of the Pa$\alpha$ survey footprint,  all six were detected via their hard X-ray emission and hence are likely massive binaries. Despite over 50 stars of comparable spectral types being found within the Arches 
(Clark et al. \cite{clark18a}, \cite{clark19a}) only four members  are known X-ray emitters (Wang et al. \cite{wang06}). While this low detection rate may in part be a function of source confusion in the crowded confines of the cluster, it raises the possibility that the `Sgr B1' cohort represents the `tip of the iceberg' in terms of a similarly rich  massive stellar population in this region.

Following on from the potential `Sgr B1 association' and, as might be anticipated given the source density in the inner reaches of the GC (Fig. 14), there are multiple  further examples of massive stars with comparatively small angular separations. Examples include (but not limited to):  P100 (O4-6 Ia$^+$) and P101 (WC8-9); P134 (WN7-8ha) and P135 (Be star); and P35 (WN8-9ha), P36 (O4-5 Ia$^+$), P111 (WN8-9ha), P112 (late-B HG/LBV)  and P114 (O4-5 Ia$^+$). The spectral classifications - and hence relative ages - of the components in the  first two  pairs disfavours a 
physical association, although this remains a possibility for the third group once P112 is excluded due to its evolutionary state. Of the remaining four stars only P35 and P114 have been subject to further examination; the radial velocity of the former  suggests that it is a runaway, while the latter appears to have formed formed  in situ (and is possibly associated with a stellar overdensity; Dong et al. \cite{dong17}). This implies that  P35 and 114 are not part of a bound physical system; although in this regard it is interesting that  P35 appears to be a member of a small group of co-moving stars, possibly the remnant of a disrupted cluster (Shahzamanian et al. \cite{shah}). Given the limitations of the current analysis no further conclusions may be drawn as to the relationship of either P36 or P111 to these stars. 

Finally, two isolated WCL stars -  2MASS J17443734-2927557 and  J17444083-2926550 - are seen in close proximity to one another 
and along the line of sight to the Sgr C H\,{\sc ii} region (RA  $\sim359^{\circ} 24' \rightarrow 359^{\circ} 28'$, $\delta \sim 0^{\circ} 4' \rightarrow -0^{\circ} 7'$; Fig. 14). The mid-IR morphology of Sgr C suggests ongoing star formation 
(Hankins et al. \cite{hankins}); if these stars are physically associated with one other and Sgr C it would extend the duration of such  activity to at least $\sim3$Myr given the time required for massive stars to progress to the WC phase (e.g. Clark et al. \cite{clark18b}). However kinematic data will be required to address this potential connection.

Indeed, it seems apparent that  due to the rich sightlines towards the GC, unambiguous  identification  of sparsely populated  clusters will require deep photometric and spectroscopic observations to identify overdensities of  
co-eval stars (cf. Sgr B), with  radial velocity and proper motion follow up needed  to  confirm the reality of putative  aggregates
(cf. the OB stars observed in the vicinity of qF353E;  Steinke et al. \cite{steinke}).
 This will be the subject of a future paper, in which we combine radial velocity measurements  from our KMOS observations with proper motion data derived from a multi-epoch HST survey of the GC (Libralato et al. \cite{libralato}).

\begin{table}
\caption{Summary of the population of massive, evolved stars within the GC 
broken down by location and spectral type.}
\begin{center}
\begin{tabular}{lccccc}
\hline
\hline
          &  Isolated & Arches & Quint. & Gal Cen. &      \\
Sub-type  &           &        &        &          & {\bf Total}\\
\hline

O4-6 Ia     &       5   &   35   &     0  &    0   & 40  \\
O7-8 Ia     &       0   &    0   &    19  &    0   & 19  \\
O9-B0 Ia    &       3   &    0   &     9  &    0   & 12  \\
OB Ia       &       1   &    0   &     10  &   26   & 37  \\
            &           &        &        &        &     \\
O4-5 Ia$^+$ &   3       &  3     &  0     &    0   &  6  \\
O5-6 Ia$^+$ &   0       &  2     &  0     &    1   &  3  \\
O6-7 Ia$^+$ &   2       &  2     &  2     &    0   &  6  \\
O7-8 Ia$^+$ &   3       &  1     &  0     &    0   &  4  \\
            &           &        &        &        &     \\
WN7-9ha     &    12     &    13  &      2 &    0   & 27  \\
            &           &        &        &        &     \\
WN9-11h/    &      9    &      0 &     10 &    8   & 27  \\
early-B HG  &           &        &        &        &     \\
            &           &        &        &        &     \\
sgB[e]/LBV/ &      8    &      0 &      3 &    0   & 11  \\       
late-B HG   &           &        &        &        &     \\
            &           &        &        &        &     \\     
WN5-7       &      8    &      0 &      1 &    4   & 13  \\ 
WN8         &      0    &      0 &      0 &    5   &  5  \\
            &           &        &        &        &     \\     
WN/WC       &   0       &      0 &      0 &    2   &  2  \\
            &           &        &        &        &     \\
WC5-6       &   0       &      0 &      0 &    1   &  1  \\
WC8-9       &   16      &      0 &     14 &   13   & 43  \\
\hline
\end{tabular}
\end{center}
{Numbers derive from this work, Clark et al. (\cite{clark18a}, \cite{clark18b}, \cite{clark19a}) and 
Paumard et al. (\cite{paumard}). We expect the census for
the Arches to be largely complete for WNLha stars and O hypergiants (Sect. 4.3); however   
following from Bartko et al. (\cite{bartko}) the numbers for
the GC cluster are lower limits, while it is almost certain that the same is true for both the 
Quintuplet cluster members and isolated stars (Sect. 4.3). Given the uncertainty in 
spectral types for supergiants  within the GC cluster we assign them a generic OB Ia 
classification; see text for details. For brevity we assign the isolated B0-3 Ia star 
 2MASS J17444501-2919307 to this cohort and the hybrid star P75 to the WN7-9ha total.
See footnotes 5 and 11 regarding the  totals derived for the 
Quintuplet; the generic OB Ia stars listed for this cluster are the faint cohort 
identified by Clark et al.
(\cite{clark18b}); $m_{\rm F205W}>13$). Finally the 35 O4-6Ia stars within the Arches include
five objects with luminosity class I-III for completeness.} 
\end{table}

\subsection{Towards a census of hot, massive stars within the GC}

We are now in a position to construct, and interpret, a census of evolved massive stars in the GC from extant data, encompassing both cluster members and isolated stars (Table 3). Given the sample size we are able to break down the supergiants, hypergiants and WRs by spectral sub-type in  order to distinguish stars of differing ages and initial masses. 
However, we explicitly exclude cool evolved stars such  as 2MASS J17444840-2902163 (Table 1), IRS 7 (Wollman et al. 
\cite{wollman}) and a number of lower luminosity stars of spectral type K-M along the line of sight to the Quintuplet ($\sim9-15M_{\odot}$; Liermann et al. \cite{liermann12}) since it is difficult to reliably distinguish these from  the foreground population of cool dwarfs and giants and hence deliver accurate population statistics. 

Observational biases associated with the census of isolated stars have  already been discussed (Sect. 4.1) but it is worth briefly addressing the limitations of current cluster surveys. Of these we expect the population of WNLha stars and O super-/hypergiants within the Arches to be essentially complete (Table 3; Clark et al. \cite{clark18a}, \cite{clark19a}); conversely the cohort  of 42 O giants  and dwarfs ($M_{\rm init}\gtrsim16M_{\odot}$) is likely to be increasingly incomplete as one moves to lower luminosities. As a consequence we omit the latter from Table 3, emphasising that our observations of the Quintuplet and the isolated stellar cohort are not sufficiently sensitive to identify such objects and hence place the respective populations in context\footnote{For this reason we also exclude the cohort of isolated classical Be stars from the census at this time (Sect. 3.7), as well as those stars of uncertain classification (Sect. 3.8). }

With current  spectral classifications only available for objects within the central $\sim40$"$\times40$" region of the more diffuse Quintuplet cluster - and being of insufficient depth to reach stars of luminosity class III-V  - the expectation is that the current census will be highly incomplete, even for very luminous sources (Table 3; Clark et al. \cite{clark19b}). This appears borne out by the low resolution and S/N spectra presented by Figer et al. (\cite{figer99}) which imply that several outlying members appear to be additional late-O/early B hypergiants of extreme intrinsic  luminosity.

Interpretation of the GC cluster is more complex for a number of reasons. Firstly, Bartko et al. (\cite{bartko}) report a further 28 WR/O stars and 34 B-dwarfs (their notation) over those described by Paumard et al. (\cite{paumard}), for a total of 177 early-type stars. Unfortunately no break down of these revised totals by spectral sub-type was presented; hence we are forced to utilise the incomplete census of the latter work in the construction of Table 3. Secondly, since there is considerable  observational uncertainty in the spectral subtypes assigned to  supergiants - ranging from $\sim$O7-B3 (Paumard et al. \cite{paumard}) -  we simply adopt a generic OB supergiant classification for the 26 stars of luminosity class I-II pending  higher S/N data. While more precise classifications are available for the WN and WC stars, modelling suggests that they are intrinsically fainter than comparable members of the  Quintuplet,
implying systematic differences in initial masses (Martins et al. \cite{martins07}, Clark et al. \cite{clark18b}); complicating any direct juxtaposition of the two samples. Finally, Paumard et al. (\cite{paumard}) identify a cohort of lower luminosity and less evolved objects within the Galactic centre cluster, reporting three O7-9 giants and 18 dwarfs stars with spectral types of O9 and later.   Following the same line of reasoning applied to the  comparable Arches cohort, we exclude these from the breakdown of stars by spectral type  presented in Table 3, as we do for the large number of additional objects of uncertain spectral type and/or luminosity class identified by these authors. 

Despite these limitations,  Table 3 lists  a total of 259 spectroscopically classified massive, evolved stars located within the GC. It is immediately apparent that isolated stars contribute significantly, comprising  $\sim26$\%  of the currently identified  population; therefore any discussion of the role of such stars in the wider ecology of the GC must account for their contribution.
This is particularly important for the cohort comprising O super-/hypergiants and WN7-9ha stars which, given their stellar properties, are likely to dominate mechanical and radiative feedback (Martins et al. \cite{martins08}, Doran et al. \cite{doran}, Lohr et al. \cite{lohr}). 

As such, it is striking that the number of WN7-9ha and O hypergiants within the  Arches cluster ($M_{\rm total}\gtrsim10^4M_{\odot}$) is directly comparable to the isolated population of such objects (Table 3). While  the current count of mid-O supergiants differs between the two settings, with only five isolated examples identified compared to 35  within the Arches, following the discussion in Sect. 4.1 we attribute this discrepancy to observational incompleteness in the field population. Likewise, comparison of the numbers of isolated early-B hypergiants, WN9-11h and WCL stars to those found within the comparably massive Quintuplet show that both populations are essentially identical in size, with the dearth of O7-8 and O9-B0 supergiants (three versus 28; Table 3) again attributable to incompleteness. 

The presence of a large number of isolated WN5-7 stars points to an additional population that is under-represented in both the Arches and Quintuplet (Table 3). Quantitative analysis of two examples - qF353 (WN6; Steinke et al. \cite{steinke}) and IRS 16SE2 (WN5/6; Martins et al. \cite{martins07}) - reveal extreme temperatures ($T_{\rm eff} >50$kK) for both stars. This suggests that, alongside the highly luminous WNLha and O-hypergiants, they may be an important source of ionising radiation in the CMZ.  We also highlight  that the population of early-B hypergiants (and WN9-11h  stars) within the GC appears unexpectedly rich when  compared to the disc population (cf. Clark et al. \cite{clark12}). Whether this is due to the particular history of the GC, with bursts of star formation occurring at times that favour the production of such stars,    observational biases disfavouring the identification of such stars in the wider galactic disc, or a more exotic explanation, such as the modification of stellar evolution due to the potentially high metallicity environment of the CMZ, is currently uncertain.  Unfortunately, the cohort of isolated sgB[e] stars, cool-phase LBVs and/or late-B hypergiants is expected to be rather  heterogeneous in terms of initial mass, thus preventing direct comparison  to the relevant subset of Quintuplet members.

Based on these number counts we suggest  that if the isolated stellar population is drawn from the same (initial) mass function as that of the Arches and Quintuplet clusters, then it should rival the combined stellar content of these aggregates. However, allowing for incompleteness and following suggestions that the quiescent star formation within the GC has proceeded at an essentially  constant rate for the past 5-10Myr ($\sim 0.1M_{\odot}$ yr$^{-1}$; Barnes et al. \cite{barnes}) it is likely that  population of isolated massive stars dominates that of the currently identified stellar clusters. Even in the case that the isolated stars are drawn from an  unexpectedly top heavy mass function - unlikely given the presence of isolated classical Be stars and low luminosity red and blue supergiants - the large number  of O super-/hypergiants, WN7-9ha and WN5-7 stars identified to date confirms  that this cohort will play an important  role in the ecology of the GC via radiative feedback.  

Nevertheless, upon consideration of all the above studies we arrive at a total  of 437 spectroscopically classified early-type/massive stars within the CMZ. These comprise 177 stars within the central cluster (Bartko et al. \cite{bartko}), 105 within the Arches (Clark et al. \cite{clark19a}), 72 within the Quintuplet (Clark et al. \cite{clark18b} and footnote 11) and 83 isolated  
examples (Sect. 3.10). 

In the absence of reliable  mass functions for both clusters and isolated stars, it is premature to employ these data to estimate the global star formation rate for the CMZ, which will be dominated by the low stellar mass component. However we may attempt to place useful lower limits to the SN rate from these number counts.  
With masses of  $\gtrsim16M_{\odot}$ expected for the O9.5 V stars within the Arches (Clark et al. \cite{clark18a}, \cite{clark19a})
and $\sim8-14M_{\odot}$ for both the isolated classical Be stars  and the B0-3 V stars within the Galactic Centre cluster (Habibi et al. \cite{habibi17}) one would expect a minimum of $\sim322$ 
($\gtrsim74$\%) of these stars to undergo core collapse within the next 20Myr (Groh et al. \cite{groh13})\footnote{Exceptions, which are limited to the GC cluster, include five B4-9 V stars as well as the large number of early-type stars of  uncertain classification.}, implying   a time-averaged  rate of $\sim2\times10^{-5}$yr$^{-1}$. More realistic assumptions lead to higher rates; for example supposing the 65 H-depleted WRs undergo core collapse in the next $\sim0.5$Myr yields a rate of $\sim1\times10^{-4}$yr$^{-1}$, with a similar number returned if we also include the mid-O super-/hypergiants, early-B hypergiants, WNLha, and WNLh stars and allow for a greater time until supernova (under the assumption that such stars derive from $M_{init}\sim40M_{\odot}$ and will undergo core-collapse within $\sim2$Myr; Groh et al. \cite{groh13}, Clark et al. \cite{clark18a}, \cite{clark18b}). 

By way of comparison, based on the number of pulsars in the GC,  Deneva et al. (\cite{deneva}) suggest a birth rate of $\gtrsim2\times10^{-4}$yr$^{-1}$, which in turn is consistent with  a SN rate of $\sim10^{-3}$yr$^{-1}$ derived from, variously, radio observations  (Lazio \& Cordes \cite{lazio}), SN remnant counts (Ponti et al. \cite{ponti}) and simulations of mass and energy flows through the innermost $\sim200$pc of the Galaxy (Crocker \cite{crocker}). While all values are comparable, we suspect the incomplete nature of the current stellar census, particularly for lower mass objects (cf. Sect. 4.1), helps to explain the order of magnitude discrepancy between the lower limit derived here and the alternative direct and indirect estimates.

\section{Conclusions and future prospects}

We have presented the results of an extensive $K-$band spectroscopic survey designed to characterise the  population of isolated 
massive stars within the GC. The resultant dataset enabled the identification of 17 new objects and the reclassification of an additional 19 known examples; a further 11 stars retained extant classifications while a large number of candidates were found to be cool, foreground sources. Including previous  identifications yields a total of 83 isolated massive stars, of which the vast majority are evolved objects.

Given the nature of the surveys utilised to construct the target list, the census is heavily biased towards objects with strong stellar winds -  such as OB hypergiants and both H-rich and H-free WRs - to the extent that even stars as extreme mid-O supergiants are significantly under-represented. Cool supergiants do not feature due to the difficulty in distinguishing  them from foreground objects. Amongst the WRs, a  large number of H-free WN5-7 stars were identified, of particular interest since they are almost entirely absent from both the Arches and Quintuplet clusters. We also report the detection of classical Be stars; given that the Be phenomenon is limited to spectral type $\sim$O9 III-V and later, these are likely to be the least massive stars ($M_{\rm init}\lesssim16M_{\odot}$) in our census. Rare transitional objects such as (candidate) sgB[e] and LBVs were also  recognised; as a consequence this population will be invaluable for constraining  massive stellar evolution - even for very rapid phases -  a possibility exemplified by P75, which appears to be the first example of an O hypergiant transitioning to a WNLha stage.  Intriguingly, a relatively large number of stars appear to be candidate binaries by virtue of their multiwavelength 
properties (Sect. 4.2.2), although the target selection criteria employed (i.e. X-ray or IR luminosity; Sect. 4.1) will preferentially identify such systems.

Isolated massive stars are found throughout the GC, although the distribution of mid-O super-/hypergiants and WNLha stars - so the youngest and most massive cohort sampled - is significantly asymmetric as a function of galactic longitude. In particular there is an overdensity of such objects spatially coincident with the young   
H\,{\sc ii} complex Sgr B1; if physically associated they would imply  star formation has proceeded in this region for at least 2Myr.
Intriguingly all these stars were identified via their hard X-ray emission; by comparison only four from fifty stars of similar spectral types within the Arches cluster are X-ray bright; on this basis one might therefore anticipate that this region hosts an exceptionally rich OB association. No further clusters comparable to the 
Arches or  Quintuplet were identifiable, despite an expectation that they should be visually prominent for at least the first  $\sim20$Myr of their lives; if such aggregates formed in the past the most  likely explanation for their current absence is tidal disruption.
 The compact stellar grouping IRS13E is likely to represent the endpoint of such a process; however, despite our survey being sensitive to such objects, no further examples were identified.

The origin of the remaining isolated massive stars is currently uncertain. A  subset appear likely  to have originally formed within a clustered environment before being  ejected; prominent candidates include the WNLha and O-hypergiants P22, 96, and 97, which are seen in close proximity to the Arches and possess identical spectral classifications to cluster members.
Nevertheless, the overall population appears too large for  them to have all originated in known clusters via such physical processes (S. Goodwin, priv. comm. 2020); hence a number were likely born  in relative isolation, possibly as the most massive component of sparse, low density aggregates. 

Comparison to extant surveys of the Arches, Quintuplet and Galactic Centre clusters allow us to place the isolated population into a wider context. Specifically, the number of isolated WNLha and O-hypergiants is directly comparable to the Arches cohort, a finding  replicated for lone WN9-11h/early-B hypergiants and WCL stars and those within the Quintuplet. 
It is currently uncertain whether the  mode(s) of star formation that generate the  high mass clusters and apparently isolated massive stars (or sparse, low mass aggregates) yield a single, universal IMF; if this is the case it would imply an isolated stellar  population that is, at a minimum,  of comparable size to that of both clusters combined. In total $\sim437$ massive cluster members and isolated  stars have been identified spectroscopically within the GC; an unprecedented number in comparison to other resolved star forming regions. Of these $\gtrsim74$\% will undergo core-collapse, 
implying a lower limit on the SN rate of $\gtrsim1\times10^{-4}$yr$^{-1}$; in reasonable agreement with other direct and indirect estimates but likely a significant underestimate given the incompleteness of the current surveys.

Moving forward, our  homogeneous, high S/N and resolution  spectroscopic dataset will permit tailored quantitative analysis for individual stars in order to derive their underlying physical parameters
(cf. Clark et al. \cite{clark18a}, \cite{clark18b}); a  significant by-product of which  will be the determination of  (differential) interstellar reddening along multiple sightlines to the GC (cf. Geballe et al. \cite{geballe}). 
Such a methodology will allow us to populate HR diagrams for both cluster and isolated cohorts. This is an essential first step in exploiting the potential of these data to improve our understanding of  massive stellar evolution as well as determining the bulk properties of the populations. Such an analysis will  enable  recent star formation across the GC to be quantitatively constrained (cf. 30 Dor; Schneider et al. \cite{schneider18}) and, critically, will permit  calibration of cluster  luminosity functions derived from  extant HST photometry in order to construct (initial) mass functions. In parallel abundance determinations - particularly of $\alpha$-elements - will help determine whether historical activity  proceeded via a mode characterised by the production of a normal or top-heavy mass function (cf. Najarro et al. \cite{paco09}). Combining the spectroscopic dataset with proper motions derived from HST  observations will allow us to identify co-moving stellar groups as well as identify high velocity massive runaways and, in turn, the relative proportions of massive stars that formed in true isolation. 

Nevertheless, additional observations are required in order to elucidate the nature and yield of recent star formation activity in the GC. The list of candidates returned by both the Paschen $\alpha$ and X-ray surveys remain to be fully exploited (cf. Appendix A), while an extension of the former to include regions such as Sgr B1 would be invaluable.
Fortuitously, despite the incomplete nature of the  current  census(es), observations of the 30 Dor star forming complex (Doran et al. \cite{doran}) suggest that   the objects identified -  and identifiable via such an approach - are expected to   dominate radiative feedback within the CMZ due to their extreme luminosities (e.g. WNLha stars and O super-/hypergiants) and/or temperatures (e.g. WN5-7 stars). However,  we are almost entirely  insensitive to the expected population of moderately massive ($\sim8-20M_{\odot}$) stars that, for any reasonable IMF, will comprise the majority of core collapse candidates. 

Photometric pre-selection followed by a proper motion  cut will be necessary to construct a suitable candidate list for exhaustive multi-object spectroscopic follow up. Although time intensive, such an approach will be required in order to determine the recent star formation rate within the GC  - and consequently the production rate of relativistic remnants - as well as the rate of feedback of mechanical energy and chemically enriched material via stellar winds and SNe. This is essential to our  understanding of the relative contributions of massive stars and the supermassive black hole Sgr A$^*$ to the energy budget and ecology of the GC, in terms of regulating the progression of star formation, the generation of very high energy (GeV) $\gamma$-ray emission, the inflation of both X-ray and radio out-of-plane bubbles centred on the GC,  and the production of PeV cosmic rays.

\begin{acknowledgements}
This research was supported by the Science and Technology Facilities Council.
F.N. acknowledges financial support through Spanish grants
ESP2017-86582-C4-1-R and PID2019-105552RB-C41
(MINECO/MCIU/AEI/FEDER) and from the Spanish State Research Agency (AEI)
through the Unidad de Excelencia “Mar\'{I}a de Maeztu”-Centro de Astrobiolog\'{I}a
(CSIC-INTA) project No. MDM-2017-0737. L.R.P. acknowledges support from the Generalitat Valenciana through the grant
PROMETEO/2019/041

\end{acknowledgements}

{}

\appendix

\section{Incomplete exploitation of the Pa$\alpha$ excess survey}

After excluding clusters members and those isolated stars with spectroscopic
observations,  27 objects from the primary list of 152 Paschen $\alpha$ excess sources of
 Dong et al. (\cite{dong11}) remain unobserved. Of these six have   colours indicative of
foreground stars  ($(H-K)<1$ - P1, 27, 55, 116, 145, and 149) while a  further
eight lack photometry (P14, 32,  33, 37 93, 138, 146, and 152). This leaves
a total of 13 stars with colours  consistent with membership of the CMZ
(P24, 26, 44, 45, 46, 51,  52, 54, 61, 76, 115, 139, and 148); however of
these only P44, 76, and 115  have $K<13$ which, excluding classical Be stars,
characterise the majority of massive stars in our survey (Sect. 3). 

The list of 189 secondary Pa$\alpha$ excess targets is much more poorly
 characterised. 101 of the isolated stars from this roster lack both 
spectroscopic and photometric observations so we may 
not  assign them to the CMZ, let alone classify them. Of the isolated stars 
currently lacking spectroscopic observations but with near-IR photometric data, 
31 appear to have colours consistent with foreground objects (S48, 50, 51, 53, 
54, 68, 70, 71, 72, 76, 77, 83, 84, 86, 89, 92, 93, 95, 97,99, 101, 109, 118, 121, 128, 145, 148, 149,
150, 154, and 170) leaving 33 possible CMZ members  (S47, 49, 52, 60, 64, 65,
78, 79, 88, 91, 103, 106, 110, 113, 114, 117, 119, 123, 125, 136, 146, 151,
153, 158, 161, 168, 169, 173, 178, 182, 183,  185, 187). Of the latter,  ten have
$K<13$ and hence are encouraging candidates for massive stars (S47, 52, 88, 103,
114, 117, 119, 125, 151, and 169).

To summarise: in comparison to our spectroscopic census,   13 stars appear to be 
strong massive star candidates,  26 fainter objects have photometric properties consistent 
with classical Be stars located within the GC, while a large cohort of 109 
stars currently lack the photometric data required to  assess of their location and nature.

\end{document}